\documentclass[10pt,a4paper]{article}

\usepackage{amsfonts,amssymb,amsmath,amsopn,amsthm,graphicx}
\numberwithin{equation}{section}
\newtheorem{theorem}{Theorem}[section]

\newtheorem{lemma}[theorem]{Lemma}
\newtheorem{corollary}[theorem]{Corollary}

\DeclareMathOperator{\curl}{curl}
\DeclareMathOperator{\directsum}{\oplus}

\DeclareMathOperator{\dist}{dist}
\DeclareMathOperator{\divergence}{div}

\DeclareMathOperator{\Laplace}{\Delta}

\DeclareMathOperator{\supp}{supp}

\def\loc{\rm{loc}}
\def\imunit{\mathrm{i}}

\newcommand{\ball}{\ensuremath{{\cal B}}}
\newcommand{\bydef}{\ensuremath{:=}}
\newcommand{\form}{\ensuremath{\mathfrak{q}}}
\newcommand{\conv}{*}
\newcommand{\D}{\ensuremath{\, d}}
\newcommand{\domain}{\ensuremath{{\cal D}}}
\newcommand{\essspectrum}{\ensuremath{\sigma_{\rm{ess}}}}
\newcommand{\ie}{i.e. }
\newcommand{\Linfty}{\ensuremath{L^{\infty}}}
\newcommand{\Ltwo}{\ensuremath{L^2}}

\newcommand{\Ordo}{\ensuremath{{\cal O}}}
\newcommand{\pdo}{\ensuremath{\partial}}
\newcommand{\real}{\ensuremath{\mathbb R}}
\newcommand{\spectrum}{\ensuremath{\sigma}}
\newcommand{\suchthat}{\ensuremath{:}}
\newcommand{\testfunc}{\ensuremath{C^{\infty}_0}}

\newcommand{\zahl}{\ensuremath{\mathbb Z}}

\title{\textbf{Stability of the magnetic Schr\"odinger operator in a waveguide}}

\author{
Tomas Ekholm \\
\small Department of Mathematics \\
\small Royal Institute of Technology \\
\small S-100 44 Stockholm, Sweden
\and
Hynek Kova\v r\'{\i}k \footnote{Also on leave of absence from Nuclear Physics
  Institute, Academy of Sciences, 25068 \v Re\v z  near Prague, Czech Republic}\\
\small Faculty of Mathematics and Physics \\
\small Stuttgart University\\
\small D-705 69 Stuttgart, Germany
}

\begin{document}

\maketitle

\begin{abstract}
The spectrum of the Schr\"odinger operator in a quantum waveguide is known to
be 
unstable in two and three dimensions. Any enlargement of the waveguide produces
eigenvalues beneath the continuous spectrum \cite{BGRS}. Also if the waveguide
is bent eigenvalues will arise below the continuous spectrum \cite{DE}. In
this paper a magnetic field is added into the system. The spectrum of the
magnetic Schr\"odinger operator is proved to be stable under small local
deformations and also under small bending of the waveguide. The proof includes
a magnetic Hardy-type inequality in the waveguide, which is interesting in its
own.
\end{abstract}

\section{Introduction}

It has been known for a long time that an appropriate bending of a two
dimensional quantum waveguide induces the existence of bound states,
\cite{ES}, \cite{GJ} and \cite{DE}. From the mathematical point of
view this means
that the Dirichlet Laplacian on a smooth asymptotical straight planar waveguide
has at least one isolated eigenvalue below the threshold of the
essential spectrum. Similar results have been obtained for a locally deformed
waveguide, which corresponds to adding a small ``bump'' to the straight
waveguide, see \cite{BGRS} and \cite{BEGK}. 
In both cases an appropriate transformation is
used to pass to a unitary equivalent operator on the straight waveguide
with an additional potential, which is proved to be attractive. As a result
at least one isolated eigenvalue appears below the essential spectrum
for {\it any} nonzero curvature, satisfying certain regularity
properties, respectively for an {\it arbitrarily small} ``bump''. The crucial
point is that for low energy the Dirichlet Laplacian in a planar waveguide in
$\real^2$ behaves effectively as a one dimensional system, in which the
Schr\"odinger operators with attractive potentials have a negative discrete
eigenvalue no matter how weak the potential is. This is related to the well
known fact that the Hardy inequality fails to hold in dimensions one and two.
\par
The purpose of this paper is to prove that in the presence of a suitable
magnetic field some critical strength of the deformation is needed for these
bound states to appear. The magnetic field is not supposed to affect
the essential spectrum of the Dirichlet Laplacian. We will deal with two
generic examples of magnetic field; a bounded differentiable field with
compact support and an Aharonov-Bohm field. The crucial technical tool of the
present work is a Hardy type inequality for magnetic Dirichlet forms in the
waveguide.  

\noindent For $d \geq 3$ the classical Hardy-inequality states that
\begin{equation} \label{firstHardy}
\int_{\real^d} \frac {|u(x)|^2}{|x|^2} \D x \ \leq \ \frac
4{(d-2)^2} \int_{\real^d} |\nabla u(x)|^2 \D x,
\end{equation}
for all $u \in H^1 (\real ^d)$. Hence if $d \geq 3$ and $V \in \testfunc
(\real^d)$,$V\geq 0$, the 
operator $-\Laplace - \varepsilon V$ does not have negative eigenvalues
for small values of the parameter $\varepsilon$.
However if $d=1,2$ then (\ref{firstHardy}) fails to hold (see
\cite{BS}) and hence the spectrum of $-\Laplace - \varepsilon V$ contains some
negative eigenvalues for any $\varepsilon>0$.
If $d=2$ and a magnetic field is introduced a higher dimensional
behavior appears. Let us consider the magnetic Schr\"odinger
operator $(-\imunit \nabla + A)^2$, where $A: \real^2 \to \real^2$
is a magnetic vector potential. In $1999$ Laptev and Weidl proved
a modified version of the inequality (\ref{firstHardy}) in $\real^2$ for the
quadratic form of a magnetic Schr\"odinger operator
\begin{equation} \label{timo2}
{\rm Const } \,\int_{\real^2} \frac {|u(x)|^2}{1+|x|^2}
\D x \ \leq \ \int_{\real^2} |(-\imunit\nabla +A)u(x)|^2 \D x,
\end{equation}
see \cite{LW}, and gave a sharp result for the case of
Aharonov-Bohm field. This was later extended in \cite{B} to
multiple Aharonov-Bohm magnetic potentials, see also \cite{EL} and
\cite{BEL}. In our model the spectrum of $(-\imunit \nabla + A)^2$ starts from
$1$ and inequality (\ref{timo2}) is not a good lower bound for functions in
$H_0^1 (\real\times(0,\pi))$. Our aim is therefore to prove that 
Hardy-inequality
\begin{equation} \label{ourinq}
{\rm Const }\,  \int_{\real\times(0,\pi)}  \frac {|u(x)|^2}{1+x_1^2} \D 
x \ \leq 
\ \int_{\real\times(0,\pi)} \big( |(-\imunit\nabla + A)u(x)|^2 - |u(x)|^2
\big) \D x, 
\end{equation}
holds true for all $u$ in the Sobolev space
$H^1_{0}(\real\times(0,\pi))$. Inequality (\ref{ourinq}) is then used to prove
stability of the spectrum of the magnetic Schr\"odinger operator under local
geometrical perturbations. 

The text is organized in the following way.
In Section 3 we prove inequality (\ref{ourinq}) for the magnetic
Schr\"odinger operator with a bounded differentiable and compactly supported field,
see Theorem \ref{generalhardytheorem}. 
The main new ingredient of our result is that we subtract the threshold of
essential spectrum. We also prove the
asymptotical behavior of the corresponding constant in the Hardy inequality in
the limit of weak fields. \par
In Section 4 we prove the stability of the essential spectrum of the operator
$(-\imunit \nabla + A)^2$ in the deformed and curved waveguide for certain magnetic
potentials, Theorem \ref{ess}. The class of magnetic potentials for which the
Theorem applies also includes the Aharonov-Bohm field.  
\par
In Section 5 we use (\ref{ourinq}) to prove that the spectrum of $(-\imunit
\nabla + A)^2$ is stable under weak deformations of the boundary of the
waveguide, Theorem \ref{existlambda}. We also give an asymptotical estimate
on the critical strength $\lambda_0$ of the deformation, for which the
discrete spectrum $(-\imunit \nabla + A)^2$ will be empty. In particular, if
the magnetic field equals $\alpha B$, then $\lambda_0$ is proportional to
$\alpha^2$ as $\alpha\to 0$. 
Moreover, we prove by a trial function argument that the same behavior of
$\lambda$, with another constant, 
is sufficient also for the presence of eigenvalues below the essential
spectrum, Theorem \ref{t5.3}. 
The latter shows that the order of $\alpha$ in our estimate is optimal. 
\par
Locally curved waveguides are studied in Section 6. We consider a waveguide
with the curvature $\beta\gamma$, where $\beta$ is a positive parameter and
$\gamma$ is some fixed smooth function with compact support. Similarly as in
Section 5 we show in  Theorem \ref{curved} that there exists a $\beta_0$, such
that for all $\beta<\beta_0$ there will be no eigenvalues in the spectrum of
$(-\imunit \nabla + A)^2$. 
The behavior of $\beta_0$ for in the limit of weak fields is at least
proportional
to $\alpha^2$, as $\alpha\to 0$. \par
The Aharonov-Bohm field requires a bit different approach due to the technical
difficulties coming from the fact that the corresponding magnetic potential
has a singularity. However, all the results mentioned above can be extended
also to this case. This is done in Section 7.

\section{The main results}

Here we formulate the main results of the paper without giving any explicit
estimates on the involved constants. For more detailed formulations see the
theorems in respectively sections.

We state the Hardy inequality for magnetic Dirichlet forms separately for the
case of an Aharonov-Bohm field and for a bounded field.

\vspace{2mm}

\noindent {\bf Theorem 3.1.} {\em Let $B \in C^1(\real^2)$ be a
bounded, real-valued magnetic field which is non-trivial in $\Omega$. Then
there is a positive constant $c$ such that
\begin{equation}
c \int_{\real \times (0,\pi)} \frac {|u| ^2}{1 + x^2} \D x \D y \ \leq \
\int_{\real \times (0,\pi)} \left( |(- \imunit \nabla + A) u|^2 - |u|^2
\right) \D x \D y, 
\end{equation}
for all $u \in H^1_0 (\real \times (0, \pi))$, where $A$ is a magnetic vector
potential associated with $B$.}

\vspace{2mm}

\noindent {\bf Theorem 7.1.} {\em
Let $A$  be the magnetic vector potential
\begin{equation}
A(x,y) = \Phi \cdot \left( \frac {- y + y_0}{x^2 + \left(y - y_0\right)},
  \frac x{x^2 + \left(y - y_0\right)}\right), 
\end{equation}
where $\Phi \in \real \setminus \zahl$ and $y_0 \in (0,\pi)$. Then there is a
positive constant $c$ such that
\begin{equation}
c \int_{\real \times (0,\pi)} \frac{|u|^2}{x^2 + (y - y_0)^2} \D x \D
y \ \leq \ \int_{\real \times (0,\pi)} \left( |(-\imunit \nabla + A) v|^2 -
  |v|^2 \right) \D x \D y, 
\end{equation}
holds for all $u \in H^1_{0,A}(\real \times (0,\pi) \setminus
    \left\{(0,y_0)\right\})$.}

\vspace{2mm}

As an application of Theorem 3.1. and Theorem 7.1. we prove
stability results for the spectrum of the magnetic Schr\"odinger
operator under geometrical perturbations. First we consider local
deformations of a waveguide. Let $f$ be a non-negative function
in $C_0^1(\real)$, $\lambda \geq 0$ and construct
\begin{equation}
\Omega_\lambda = \left\{(x,y) \in \real^2 \suchthat 0 < y < \pi +
\lambda \pi f(x)\right\}.
\end{equation}
Let $M_d$ be the Friedrich's extension of the operator
\begin{equation}
\left( - \imunit \pdo_x + a_1 \right)^2 + \left( - \imunit \pdo_y
+ a_2 \right)^2,
\end{equation}
defined on $\testfunc (\Omega_\lambda)$, where $A$ is either the
magnetic vector potential for the Aharonov-Bohm field inside the
waveguide or a magnetic vector potential associated with a
magnetic field $B \in C_0^1(\real^2)$, such that $B$ is
non-trivial in $\Omega_\lambda$. Then the following statement
holds:

\vspace{2mm}

\noindent {\bf Theorem 5.1. and 7.4.} {\em There is a positive constant
  $\lambda_0$ such that for $\lambda \in (0, \lambda_0)$ the operator $M_d$
  has purely essential spectrum $[1, \infty)$.}

\vspace{2mm}

Assume that we replace the field $B$ by $\alpha B$, where $\alpha > 0$ then
there are constants $c_a$ and $c_e$ such that if
\begin{equation}
\lambda < c_a \alpha^2 + \Ordo (\alpha^4),
\end{equation}
as $\alpha \to 0$, then the discrete spectrum of $M_d$ is empty. But if
\begin{equation}
\alpha^2 < c_e \lambda + \Ordo (\lambda^2),
\end{equation}    
as $\lambda \to 0$, then $M_d$ has at least one eigenvalue.

If we now consider $M_c$ being the same operator as $M_d$ but in a curved
waveguide $\Omega_\beta$, where $\beta\gamma$ indicates the curvature of the
boundary of the waveguide the results are similar.

\vspace{2mm}

\noindent {\bf Theorem 6.1. and 7.5.} {\em There is a positive constant
  $\beta_0$ such that if $\beta \in (0, \beta_0)$ then the operator $M_c$ has
  purely essential spectrum $[1, \infty)$.} 

\section{A Hardy-type inequality} \label{chaptergen}

In this section we will prove a Hardy inequality in the case of a
general bounded, differentiable magnetic field.

Let $\Omega = \real \times (0, \pi)$ and let $B$ be a bounded,
real-valued magnetic field such that $B \in C^1(\real^2)$ and $B$
is non-trivial in $\Omega$. Choose a point $p \in \Omega$ such
that there is a ball $\ball_R(p) \subset \Omega$ with
\begin{equation}\label{B}
\Phi(r) \bydef \frac{1}{2\pi} \int_{\ball_r(p)} B(x,y) \D x \D y
\end{equation}
not identically zero for $r \in (0,R)$. For simplicity let $p =
(0,y_0)$, for some $y_0 \in (0, \pi)$.

We can construct a magnetic vector potential for $B$ as
  $A(x, y) = (a_1(x, y), a_2(x, y))$ defined on $\real^2$ in the
  following way
\begin{eqnarray}
a_1 (x, y) &=& - (y - y_0) \int_0^1 B(u x, u(y - y_0) + y_0) \, u
\D u, \label{a_1} \\ a_2 (x, y) &=& x \int_0^1 B(u x, u(y - y_0) +
y_0) \, u \D u. \label{a_2}
\end{eqnarray}
Then $(\curl A)(x,y) = \partial_x a_2(x,y) - \partial_y a_1(x,y) =
B(x,y)$ and the transversal gauge $A(x,y) \cdot (x, y - y_0) = 0$
for all $(x,y) \in \real^2$ is satisfied. Note that since
$a_1,a_2 \in \Linfty (\real^2)$ we have $H_{0,A}^1(\Omega) =
H_0^1(\Omega)$, where $H_{0,A}^1(\Omega)$ denotes the completion of $\testfunc
(\Omega)$ in the norm
\begin{equation}
\|u\|_{H_{0,A}^1(\Omega)}^2 = \|u\|_{\Ltwo (\Omega)}^2 + \|(-\imunit \nabla +
A)u\|_{\Ltwo (\Omega)}^2.
\end{equation}

\begin{theorem} \label{generalhardytheorem}
Let $B \in C^1(\real^2)$ be a real-valued magnetic field such that
$B \not\equiv 0$ in $\Omega$. Then 
\begin{equation} \label{generalhardyinequality}
c_H \int_\Omega \frac {|u|^2}{1 + x^2} \D x \D y \leq \int_\Omega
\left( |(- \imunit \nabla + A) u|^2 - |u|^2 \right) \D x \D y,
\end{equation}
holds for all $u \in H_0^1 (\Omega)$, where $A$ is a magnetic vector
potential associated with $B$ and $c_H$ is a positive constant, given
in {\rm (\ref{finalconst})}. 
\end{theorem}

\begin{proof}
Due to gauge invariance of the inequality
(\ref{generalhardyinequality}) we can without loss of generality
assume that the components of $A$ are given by (\ref{a_1}) and
(\ref{a_2}). Let $(r, \theta)$ be polar coordinates centered at the
point $p$. We will prove that the inequality
\begin{equation}
c \int_{\ball_R(p)} |u|^2 r \D r \D \theta \leq \int_{\ball_R(p)}
\left( |u_r|^2 + r^{-2} |-\imunit u_\theta + r
    a(r,\theta) u|^2\right) r \D r \D \theta,
\end{equation}
holds for all $u \in H^1_0(\Omega)$, where $a(r,\theta) = A \cdot
(-\sin \theta , \cos \theta)$ and $c$ is a positive constant.

For fixed $r$ we consider the operator $K_r = -\imunit
\partial_\theta + r a(r,\theta)$ in $\Ltwo (0,2\pi)$, which was
studied in \cite{LW}. The operator $K_r$ is self-adjoint on the
domain $H^1 (0,2\pi)$ with periodic boundary conditions. The
spectrum of $K_r$ is discrete and the eigenvalues
$\left\{\lambda_k\right\}_{k = - \infty}^\infty$ and the
orthonormal set of eigenfunctions $\left\{\varphi_k\right\}_{k = -
\infty}^\infty$ are given by
\begin{equation}
\lambda_k = \lambda_k (r) = k + \frac r{2\pi}
\int_0^{2\pi}a(r,\theta) \D \theta = k + \Phi(r),
\end{equation}
and
\begin{equation}
\varphi_k (r,\theta) = \frac 1 {\sqrt{2\pi}} e^{\imunit \lambda_k
\theta - \imunit r \int_0^\theta a(r,s) \D s}.
\end{equation}
The quadratic form of $K_r^2$ satisfies the following inequality
\begin{equation}
\mu(r)^2 \int_0^{2\pi} |u|^2 \D \theta \leq \int_0^{2\pi}
|-\imunit u_\theta + r a u|^2 \D \theta,
\end{equation}
for all $u(r,\cdot) \in H^1 (0,2\pi)$, where $\mu(r) = \dist
(\Phi(r), \zahl)$. Thus
\begin{equation}
\int_{\ball_R(p)} \frac{\mu^2}{r^2}|u|^2 r \D r \D \theta \leq
\int_{\ball_R(p)} r^{-2}|-\imunit u_\theta + r a u|^2 r \D r \D
\theta,
\end{equation}
holds for all $u \in H^1 (\Omega)$.

Define the function $\chi : [0,R] \rightarrow [0,1]$ by
\begin{equation}
\chi(r) = \frac{\mu_0^2 \, \mu(r)^2}{r^2}, \quad \text{ where
}\quad \mu_0 = \left(\max_{r \in [0,R]}
\frac{\mu(r)}{r}\right)^{-1}.
\end{equation}
Since $\Phi$ is piecewise continuous differentiable and $\Phi(0) =
0$ it follows that $\chi$ is well defined. It is clear that
$\chi(r) \in [0,1]$ and that there exists at least one $r_0 \in
(0,R]$ such that $\chi(r_0) = 1$.
Let $v \in H^1(0,R)$ such that $v(r_0) = 0$, then we have the
following inequalities
\begin{equation} \label{lapcir}
\int_{r_0}^R |v(r)|^2 r \D r \leq \frac {2R^3 - 3 R^2 r_0 +
r_0^3}{6r_0} \int_{r_0}^R |v'(r)|^2 r \D r,
\end{equation}
and
\begin{equation} \label{fromLW}
\int_0^{r_0} |v(r)|^2 r \D r \leq \frac {r_0^2}{\nu_0^2}
\int_0^{r_0} |v'(r)|^2 r \D r,
\end{equation}
where $\nu_0 \geq 2$ is the first zero of the Bessel function
$J_0$. The latter comes from the lowest eigenvalue of $-\Delta$ in
a circle with Dirichlet boundary conditions at the radius $r_0$.
The first inequality follows by writing
\begin{equation}
|v(r)|^2 = \left| \int_{r_0}^r v'(t) \D t \right|^2 \leq (r - r_0)
\int_{r_0}^R |v'(r)|^2 \D r.
\end{equation}
Using (\ref{lapcir}) and (\ref{fromLW}) we conclude that
\begin{eqnarray} \label{gen2}
\int_{\ball_R(p)} |u|^2 r \D r \D \theta &\leq& 2
\int_{\ball_R(p)} \left( |\chi
  u|^2 + |(1 - \chi)u|^2 \right) r \D r \D \theta \nonumber \\
&\leq& 2 \mu_0^2 \int_{\ball_R(p)} r^{-2} |-\imunit u_\theta + r a u|^2 r \D r
  \D   \theta \nonumber \\
&& + \, 2 \int_0^{2\pi} \bigg( \frac{r_0^2}{\nu_0^2} \int_0^{r_0} |((1-\chi)u)'|^2
  r \D r \nonumber \\
&& + \frac {2R^3 - 3 R^2 r_0 + r_0^3}{6r_0} \int_{r_0}^R |((1 - \chi)u)'|^2 r
  \D r \bigg) \D \theta \\
&\leq& 2 \mu_0^2 \int_{\ball_R(p)} r^{-2} |-\imunit u_\theta + r a u|^2 r \D r
  \D   \theta \nonumber \\
&& + \, c_0 \int_{\ball_R(p)} \left( | \chi' u |^2 + | u_r |^2
\right) r \D r \D \theta \nonumber \\
&\leq& c_1 \int_{\ball_R(p)} \left( |u_r|^2 + r^{-2} |-\imunit u_\theta +
  r a u|^2 \right) r \D r \D \theta, \nonumber
\end{eqnarray}
where
\begin{eqnarray}
c_0 &=& 4 \max \left\{ \nu_0^{-2} r_0^2, (6r_0)^{-1}(2R^3 - 3 R^2
r_0 + r_0^3) \right\}, \\ c_1 &=& \max \left\{ 2 \mu_0^2 + 4 c_0
c_2^2 \mu_0^4, c_0 \right\} , \\ c_2 &=& \max_{r \in [0,R]}
|r^{-2}(r \mu'(r) - \mu(r))|.
\end{eqnarray}

The operator $-\frac {d^2}{dy^2} - 1$ on the domain $\left\{ u \in
  H_0^2(0,\pi) \suchthat u(y_0) = 0\right\}$ is greater or equal to
\begin{equation}
c_3 \bydef \pi^2 \min \left\{ y_0^{-2}, (\pi - y_0)^{-2}\right\} -
1.
\end{equation}
This can be easily verified by writing $-\frac {d^2}{dy^2} - 1$ as
the direct
  sum $\left( - \frac {d^2}{dy^2} - 1\right) \directsum \left(- \frac
  {d^2}{dy^2} - 1 \right)$ on the set
  $H_0^2(0,y_0) \directsum H_0^2(y_0,\pi)$. In terms of quadratic
  forms this means that for $v$ in $H^1 (0,\pi)$ we have
\begin{equation}\label{316}
\int_0^\pi |v(y)|^2\sin^2 y \D y \leq c_3^{-1} \int_0^\pi
|v'(y)|^2 \sin^2 y \D y.
\end{equation}
Let $u \in H^1(\Omega)$ and $\psi : (0,\pi) \to [0,1]$ be defined by
\begin{equation} \label{smallpsi}
\psi(y) = \left\{
\begin{array}{c@{\quad,}l}
\frac{|y - y_0|}{\sqrt{R^2 - x^2}} & \text{ if } h_-(x) < y < h_+(x), \cr
1 & \text{ otherwise.}
\end{array}
\right. 
\end{equation}
where $h_\pm (x) = y_0 \pm \sqrt{R^2 - x^2}$. We write $u = (1 -
\psi)u + \psi u$ and use (\ref{316}) to obtain
\begin{eqnarray} \label{319}
\int_0^\pi |u|^2 \sin ^2 y \D y &\leq& 2
\int_{h_-(x)}^{h_+(x)} |(1-\psi)u|^2 \sin^2 y \D y \\
&& + \frac 4{c_3} \left( \int_0^\pi |u_y \psi|^2 \sin^2 y \D y +
  \int_{h_-(x)}^{h_+(x)} \frac {|u|^2 \sin^2 y \D y}{R^2 - x^2} \right)
\nonumber .
\end{eqnarray}
Let $\Omega_R = (-R,R) \times (0,\pi)$, then by (\ref{gen2}) and
(\ref{319}) we get
\begin{eqnarray}
\int_{\Omega_R} (R^2 - x^2) |u|^2 \sin^2 y \D y \D x &\leq&
\frac{c_1 (2 R^2 c_3 + 4)}{c_3 \cos^2 (|y_0 - \frac \pi 2| + R)}
 \int_{\ball_R(p)} |(-\imunit\nabla + A)u|^2\, \sin^2 y  \D x \D y \nonumber \\
&& + \frac{4R^2}{c_3}\, \int_{\Omega_R} |u_y|^2 \sin^2 y \D
y \D x ,
\end{eqnarray}
for all $u \in H^1(\Omega)$. If $u = |v|$ where $v \in C^\infty
(\overline \Omega)$ then by the diamagnetic inequality (see for instance
\cite{K}, \cite{S}, \cite{AHS} and \cite{HS}) saying that
\begin{equation} \label{diamagneticinequality}
|\nabla |v|(x,y)| \leq |(-\imunit \nabla + A)v(x,y)|
\end{equation}
holds almost everywhere, it follows that
\begin{equation}\label{321}
\int_{\Omega_R} (R^2 - x^2) |u|^2 \sin^2 y \D x \D y \leq c_4
\int_{\Omega_R} |(- \imunit \nabla + A) u|^2 \sin^2 y \D x \D
y ,
\end{equation}
holds for all $u \in C^\infty (\overline \Omega)$ with
\begin{equation}
c_4= \frac{2R^2 c_1 c_3 + 4c_1 + 4R^2}{ c_3 \, \cos^2(|y_0 -
\frac{\pi}{2}| + R)} \, .
\end{equation}
We need the classical one-dimensional Hardy inequality saying that
\begin{equation}\label{classicalhardy}
\int_{-\infty}^{\infty} \frac{|v|^2}{t^2} \D t \ \leq \ 4
\int_{-\infty}^{\infty} |v'|^2 \D t,
\end{equation}
holds for any $v \in H^1 (\real)$, such that $v(0)=0$ (see
\cite{H}).
Take $m = \frac R{\sqrt 2}$ and let the mapping $\varphi : \real
\to [0,1]$ be defined by
\begin{equation} \label{smallphi}
\varphi(x) \bydef \left\{
\begin{array}{l@{\quad, \quad}l}
1 & \text{if } |x| > m, \\
\frac {|x|}m & \text{if } |x| < m.
\end{array}
\right.
\end{equation}
Let $u \in C^\infty (\overline \Omega) \cap \Ltwo (\Omega)$, by writing $u =
u\varphi + u (1 - \varphi)$ and using (\ref{diamagneticinequality}),
(\ref{321}) and (\ref{classicalhardy}) we obtain
\begin{eqnarray}
\int_\Omega \frac {|u|^2 \sin^2 y}{1 + x^2} \D x \D y &\leq& 2
\int_\Omega \frac{|u \varphi|^2 + |u (1 - \varphi)|^2}{1
+ x^2} \sin^2 y \D x \D y \\
&\leq& 16 \int_\Omega \left( |u_x \varphi|^2 + |u \varphi'|^2 \right)
\sin^2 y \D x \D y + 2 \int_{\Omega_m} \frac{|u|^2 \sin^2
y}{1 + x^2} \D x \D y \nonumber \\
&\leq& 16 \int_\Omega |u_x| ^2 \sin^2 y \D x \D y + c_5
\int_{\Omega_R} (R^2 - x^2) |u|^2 \sin^2 y \D x \D y \nonumber\\
&\leq& c_6 \int_\Omega |(- \imunit \nabla + A) u|^2 \sin^2 y \D x \D
y, \nonumber
\end{eqnarray}
where
\begin{equation} \label{c_5}
c_5 = \frac{64 + 4R^2}{R^4} \qquad \text{ and } \qquad c_6 = 16 +
c_4 c_5.
\end{equation}
If we now substitute $v(x,y) = u(x,y) \sin y$ the statement of the
theorem with 
\begin{equation} \label{finalconst}
c_H = c_6^{-1}
\end{equation}
will follow by continuity.
\end{proof}
Let us replace the field $B$ by $\alpha B$, where $\alpha$ is a positive
constant. Let $\Phi_B$ be defined by (\ref{B}) with the field $B$
and define the following constants.
\begin{eqnarray}
k_1 &=& \left(\max_{r \in [0,R]} r^{-1} \Phi_B (r)\right)^{-1}, \\
k_2 &=& \max_{r \in [0,R]} |r^{-2}(r \Phi_B'(r) - \Phi_B(r))|, \\ k_4
&=& \frac{(2R^2c_3 + 4)(2k_1^2 + 4 c_0 k_1^4 k_2^2)}{c_3
\cos^2(|y_0 - \frac{\pi}{2}| + R)}. \label{k_4}
\end{eqnarray}

\begin{corollary} \label{ascor}
If we replace $B$ by $\alpha B$ in Theorem {\rm
\ref{generalhardytheorem}}, then the constant $c_H$ in {\rm
(\ref{generalhardyinequality})} satisfies the following equality
\begin{equation} \label{generalhardyconstant}
c_H \geq \frac{1}{k_4c_5} \, \alpha^2 + \Ordo (\alpha^4),
\end{equation}
for $\alpha \to 0$.
\end{corollary}

\begin{proof}
We first note that the constants $c_0$, $c_3$ and $c_5$ are
independent of $\alpha$. As $\alpha \to 0$ the constant $c_1 =
(2k_1^2 + 4 c_0 k_1^4 k_2^2) \alpha^{-2}$ and $c_2 = k_2 \alpha$.
This implies that $c_4 = k_4 \alpha^{-2} + \Ordo (1)$ and
therefore (\ref{generalhardyconstant}) holds as $\alpha \to 0$.
\end{proof}

\section{Stability of essential spectrum}

Let $\Omega$ be a subset of $\real^2$ with $\partial \Omega$ being
piecewise continuously differentiable and let us assume that there
is a bounded set $\Omega_0 \subset \real^2$ such that $\Omega
\setminus \Omega_0$ consists up to translations and rotations of
two half strips $\Omega_1$ and $\Omega_2$. By a half strip we
denote the set $(0, \infty) \times (0, \pi) \setminus P$, where
$P$ is either empty or contains finite number of points in $\real^2$.
Let $M$ be the operator $(- \imunit \nabla + A)^2$ on $H_{0,A}^2 (\Omega)$,
for some magnetic vector potential $A$.
\begin{theorem} \label{ess}
If the magnetic vector potential $A = (a_1, a_2)$ is such that for $j = 1,2$
we have $a_j \in \Ltwo_{\loc} (\Omega_1 \cup \Omega_2)$, $a_j \in L^{2 +
  \varepsilon} (\Omega_0)$ for some $\varepsilon > 0$ and the functions $|A|$
and $\divergence A$ are for some $R > 0$ in $\Ltwo \left(\Omega_2 \cap \left\{
  x \in \real^2 \suchthat |x| > R \right\}\right)$, then
\begin{equation}
\essspectrum (M) = [1,\infty).
\end{equation}
\end{theorem}
\begin{proof}
We can without loss of generality assume that $\Omega_2 =
(0,\infty) \times (0, \pi)$. To prove that $[1,\infty) \subset
\essspectrum (M)$ we construct Weyl sequences. Assume that
$\lambda$ is a non-negative real number. Let
$\left\{h_n\right\}_{n = 1}^\infty$ be a singular sequence of
real-valued testfunctions for the operator $-\frac{d^2}{dx^2}$ in
$\Ltwo (\real)$ at $\lambda$ such that $\supp
  h_n \in (n,\infty)$ and such that $\|h_n\|_\infty$ and
  $\|h'_n\|_\infty$ are uniformly bounded in $n$. For
  instance let $\varphi \in \testfunc (\real)$ be a non-negative
  function such that $\|\varphi\|_{\Ltwo (\real)} = 1$ and $\supp
  \varphi \subset (-1,1)$. Let
\begin{equation}
\rho_n (x) = \left\{ \begin{array}{l@{\quad , \text{ if} \quad}l}
0 & x < n \text{ or } x \geq n^2, \\ \frac {2x}{n(n-1)} - \frac
2{n-1} & n \leq x < \frac{n(n+1)}{2}, \\ \frac {-2x}{n(n-1)} +
\frac {2n}{n-1} & \frac{n(n+1)}{2} \leq x < n^2,
              \end{array}
\right.
\end{equation}
then $h_n$ can be chosen as a subsequence of $(\rho_n \conv
\varphi)(x) \cdot \cos (\sqrt \alpha x)$ such that the functions
from the subsequence have disjoint support.

Construct the functions
\begin{equation}
g_n(x,y) = h_n(x) \sin y.
\end{equation}
We will prove that $g_n$ is a singular sequence for $M$ at $1 +
\lambda$. Clearly $g_n \in \domain (M)$ for $n$ large enough and
\begin{equation}
\|g_n\|_{\Ltwo (\Omega)}^2 = \int_{\Omega} h_n(x)^2 \sin^2 y \D x
\D y = \frac {\pi}2 \|h_n\|_{\Ltwo (\real)}^2 > 0,
\end{equation}
for every $n$.

Let $u$ be any function in $\Ltwo (\Omega)$, then
\begin{eqnarray}
(u, g_n)_{\Ltwo (\Omega)} = \int_0^{\pi} \sin y
  \int_n^{\infty} h_n(x) u(x,y) \D x \D y \to 0,
\end{eqnarray}
the latter follows since $u(\cdot, y)$ is in $\Ltwo (\real)$ for
a.e. $y \in (0, \pi)$. Finally we must show that $(M - (\lambda +
1))g_n \rightarrow 0$, as $n \rightarrow \infty$. There is a
constant $c$ depending on $\|h_n\|_\infty$ and $\|h'_n\|_\infty$
such that

\begin{align}
\|(M - (1 + \lambda))g_n\|_{\Ltwo (\Omega)}^2 &= c \bigg(
\int_n^\infty | - h_n' - \lambda h_n|^2 D x \\
& \quad \quad + \int_0^\pi\int_n^\infty \left( |A|^2 + |\divergence A|^2
\right) \D x \D y \bigg) \to 0,
\end{align}
as $n \rightarrow \infty$. We have proved that $1 + \lambda \in
\essspectrum (M)$ for all non-negative $\lambda$, \ie $[1,\infty)
\subset \essspectrum (M)$.

To prove the reverse inclusion $\essspectrum (M) \subset
[1,\infty)$ it will be enough to prove that $\inf \essspectrum (M)
\geq 1$. We study the operator $M_N$ being $M$ with additional
Neumann boundary condition at the intersections $\Omega_0 \cap
\Omega_1$ and $\Omega_0 \cap \Omega_2$. Then $M_N$ can be written
as a direct sum of three operators $M_1 \directsum M_0 \directsum
M_2$ on the domain $H_{0,A}^2(\Omega_1) \directsum
H_{0,A}^2(\Omega_0) \directsum H_{0,A}^2(\Omega_2)$. Since the magnetic field
is in $L^{2 + \varepsilon}(\Omega_0)$ the norms in $H^1_A(\Omega_0)$ and
$H^1_0(\Omega_0)$ are equivalent. This implies that the spectrum of $M_0$ is
discrete. By the maximin principle we have
\begin{equation}
\inf \essspectrum (M) \geq \inf \essspectrum (M_N) = \inf
\essspectrum (M_2) \geq \inf \spectrum (M_2).
\end{equation} 
By the diamagnetic inequality we get that
\begin{equation}
\inf \spectrum (M_2) \geq \inf \spectrum (- \Delta) = 1.
\end{equation}
The last inequality follows since Dirichlet boundary conditions in the points
contained in $P$ don't affect the spectrum of $-\Delta$. Hence the proof is 
complete. 
\end{proof}

\section{Locally deformed waveguides}

Let $f$ be a non-negative function in $C^1_0 (\real)$ and for
$\lambda \geq 0$ we construct
\begin{equation}
\Omega_\lambda = \left\{(s,t) \in \real^2 \suchthat 0 < t < \pi +
\lambda \pi f(s)\right\}.
\end{equation}
In \cite{BGRS} it was proven that the Friedrich's extension of
$-\Delta-1$ defined on $\testfunc (\Omega_\lambda)$ had negative
eigenvalues for all $\lambda > 0$. For small enough values of
$\lambda>0$ there is a unique simple negative eigenvalue
$E(\lambda)$, the function $E(\lambda)$ is analytic at $\lambda =
0$ and
\begin{equation}
E(\lambda) = - \lambda^2 \left(\int_\real f(s) \D s\right)^2 +
\Ordo (\lambda^3).
\end{equation}
We will show that if we add a magnetic field to the Schr\"odinger
operator it will prevent these negative eigenvalues to appear for
small values of $\lambda$.

Assume that $B \in C_0^1(\real^2)$ such that $B$ is not
identically zero in $\Omega_\lambda$. Let $M_d$ be the Friedrich's
extension of the symmetric, semi-bounded operator
\begin{equation}
\left( -\imunit \pdo_s + a_1(s,t)\right)^2 + \left( -\imunit
\pdo_t + a_2(s,t)\right)^2,
\end{equation}
defined on the domain $\testfunc (\Omega_\lambda)$, where $A(s,t)
= (a_1(s,t), a_2(s,t))$ is a magnetic vector potential associated
with $B$. Due to gauge invariance we can assume that $A$ is
defined by (\ref{a_1}) and (\ref{a_2}). Since $B$ is
bounded and of compact support, it follows from (\ref{a_1}) and
(\ref{a_2}) that $a_1, a_2 \in \Linfty (\real^2)$ and for $r =
|(s,t)| \to \infty$ we have
\begin{equation} \label{Aisdecreasing}
|a_j(s,t)| = \Ordo (r^{-1}), \text{ for } j=1,2.
\end{equation}
This implies that the essential spectrum of $M_d$ coincides by Theorem
\ref{ess} with the half-line $[1,\infty)$.

\begin{theorem} \label{existlambda}
There is a positive number $\lambda_0$ depending on
$\|f\|_\infty$, $\|f'\|_\infty$, $\|a_1\|_\infty$ and
$\|a_2\|_\infty$ such that for $\lambda \in (0, \lambda_0)$ the
discrete spectrum of $M_d$ is empty.
\end{theorem}
\begin{proof}
We denote by $\form_d$ the quadratic form associated with $M_d$, \ie
\begin{equation}
\form_d [\psi] = \int_{\Omega_\lambda} \left(| - \imunit \psi_s +
a_1 \psi|^2 + | - \imunit \psi_t + a_2 \psi|^2\right) \D s \D t,
\end{equation}
with $\domain (\form_d) = H_0^1(\Omega_\lambda)$. Define
\begin{equation} \label{u1}
U_\lambda : \Ltwo (\Omega_\lambda) \to \Ltwo (\Omega_0)
\end{equation}
to be the unitary operator given by
\begin{equation} \label{u2}
(U_{\lambda} \psi) (x,y) = \sqrt{1+\lambda f(x)} \psi (x, (1 +
  \lambda f(x))y).
\end{equation}
The operator $M_d$ is unitary equivalent to the operator
\begin{equation} \label{413}
M_\lambda \bydef U_{\lambda} M_d U^{-1}_{\lambda},
\end{equation}
defined on the set $U_{\lambda}\domain (M_d)$ in $\Ltwo
(\Omega_0)$. The form associated with $M_\lambda$ is then given by
\begin{equation}
\form_\lambda [\varphi] = \form_d [U_\lambda^{-1} \varphi],
\end{equation}
defined on the space $\domain (\form_\lambda) = U_\lambda \domain
(\form_d)$. If we prove that $M_\lambda - 1$ is non-negative, then
the theorem will follow from (\ref{413}) and the fact that
$\essspectrum (M_d) = [1, \infty)$.

For convenience let $g(s) = 1 + \lambda f(s)$, then
\begin{eqnarray}
\form_\lambda[\varphi] &=& \form_d[U^{-1}_\lambda\varphi]
\nonumber \\
&=& \int_{\Omega_{\lambda}} \left( \left| (-\imunit \pdo_s + a_1(s,t))
(g(s)^{-\frac 12} \varphi (s, g(s)^{-1}t)) \right|^2 \right.
\nonumber \\
&& \qquad \left. + \left| (-\imunit \pdo_t + a_2(s,t)) (g(s)^{-\frac 12}
\varphi (s, g(s)^{-1}t)) \right|^2 \right) \D s \D t \nonumber \\
&=& \int_{\Omega_0} \left( \bigg| \frac {\imunit g'(x)}{2g(x)} \varphi(x,y) -
\imunit {\varphi}_x (x,y) \right. \\
&& \qquad \quad + \frac{\imunit y g'(x)}{g(x)} \varphi_y (x,y) +
\tilde a_1(x,y)\varphi(x,y) \bigg|^2 \nonumber \\
&& \qquad \left. \quad + \left|-\frac {\imunit}{g(x)} \varphi_y (x,y) +
\tilde a_2(x,y)\varphi(x,y) \right|^2 \right) \D x \D y \nonumber
,
\end{eqnarray}
where
\begin{equation} \label{modAB}
\tilde A(x,y) = (\tilde a_1(x,y), \tilde a_2(x,y)) = A(x, g(x)y).
\end{equation}
Straightforward calculation gives
\begin{eqnarray} \label{416}
\form_\lambda[\varphi]
&=& \int_{\Omega_0} \bigg( \left| -\imunit \varphi_x + \tilde a_1 \varphi
\right|^2 + \left| -\imunit \varphi_y + \tilde a_2 \varphi
\right|^2 - |\varphi_y|^2 \nonumber \\
&& \quad -\frac{g'}{2g}
   (\varphi\overline{\varphi_x} + \varphi_x \overline{\varphi})
- \frac 14 \left(\frac{g'}{g} \right)^2|\varphi|^2 - \frac{y g'}g
(\varphi_x \overline{\varphi_y} + \varphi_y \overline{\varphi_x})
\\
&& \quad + \frac{y^2(g')^2 + 1}{g^2}|\varphi_y|^2
+ \imunit \frac{y g' \tilde a_1 + \lambda f \tilde a_2}g
(\varphi_y \overline{\varphi} - \varphi\overline{\varphi_y})
\bigg) \D x \D y \nonumber .
\end{eqnarray}
Let $\form$ be the quadratic form associated with the
Schr\"odinger operator with the magnetic vector potential $\tilde
A$ in the space $\Ltwo (\Omega_0)$. We have
\begin{eqnarray} \label{generaldeform}
\form_\lambda [\varphi] - \|\varphi\|_{\Ltwo (\Omega_0)}^2
&=& \form [\varphi] - \|\varphi\|_{\Ltwo (\Omega_0)}^2
\nonumber \\
&& + \int_{\Omega_0} \left( \frac{y^2 \lambda^2 (f')^2 - 2 \lambda
  f - \lambda^2 f^2}{g^2} |\varphi_y|^2 - \frac 14 \left(\frac{\lambda
    f'}g \right)^2 |\varphi|^2 \right. \nonumber \\
&& \qquad - \frac{y \lambda f'}g (\varphi_x \overline {\varphi_y} +
\varphi_y \overline{\varphi_x}) - \frac{\lambda f'}{2g} (\varphi
\overline{\varphi_x} + \varphi_x \overline \varphi) \\
&& \qquad \left. + \imunit \lambda \frac{ y f' \tilde a_1 + f \tilde
  a_2}g (\varphi_y \overline \varphi - \varphi
  \overline{\varphi_y}) \right) \D x \D y. \nonumber
\end{eqnarray}
Without loss of generality we can assume that $\lambda \leq 1$. Let $\chi$ be
the characteristic function of the support of $f$. The following lower bound
holds true,
\begin{equation}
\form_\lambda [\varphi] - \|\varphi\|_{\Ltwo (\Omega_0)}^2 \geq
\form [\varphi] - \|\varphi\|_{\Ltwo (\Omega_0)}^2 - \lambda
\int_{\Omega_0} \chi \cdot \left( c_7 \left( |\varphi_x|^2 +
|\varphi_y|^2 \right) + c_8 |\varphi|^2 \right) \D x \D y,
\end{equation}
where the constants are given by
\begin{eqnarray}
c_7 &=& \|f\|^2_\infty + (2 + \|a_2\|_\infty) \|f\|_\infty +
(2^{-1} + \pi + \pi \|a_1\|_\infty) \|f'\|_\infty, \\ c_8 &=&
4^{-1} \|f'\|_\infty^2 + 2^{-1} \|f'\|_\infty + \pi 
\|a_1\|_\infty \|f'\|_\infty + \|a_2\|_\infty \|f\|_\infty.
\end{eqnarray}
By the pointwise inequality
\begin{equation}\label{abtoa2b2}
|\varphi_x|^2 + |\varphi_y|^2 \leq 2\left(| - \imunit \nabla
\varphi + \tilde A\varphi|^2 +  |\tilde A|^2 |\varphi|^2 \right)
\end{equation}
and Theorem \ref{generalhardytheorem} we get
\begin{eqnarray} \label{sistaG}
\form_\lambda [\varphi] - \|\varphi\|_{\Ltwo (\Omega_0)}^2
&\geq& \left( \frac 12 - 2 \lambda c_7 \right) \left( \form
[\varphi] - \|\varphi\|_{\Ltwo (\Omega_0)}^2 \right) \\
&& + \left( \frac {c_H}2 - \lambda c_9 (1 + d^2) \right) \int_{\Omega_0}
\frac {|\varphi|^2}{1 + x^2} \D x \D y, \nonumber
\end{eqnarray}
where
\begin{equation}\label{dandc_9}
d = \max \supp f \quad \text{ and } \quad c_9 = 2(1 +
\|a_1\|_\infty^2 + \|a_2\|_\infty^2)c_7 + c_8
\end{equation}
and $c_H$ is the constant from (\ref{generalhardyinequality}). Let
\begin{equation}
\lambda_0 = \frac {c_H}{2c_9(1 + d^2)},
\end{equation}
then the right hand side of (\ref{sistaG}) is positive for all
$\lambda \in (0, \lambda_0)$.
\end{proof}

If we replace $B$ by $\alpha B$, $A$ will be replaced $\alpha A$. Let us
define
\begin{equation}
k_9 := \lim_{\alpha\to 0} c_9 = \|f\|_\infty^2 +
  2\|f\|_\infty +4^{-1}\|f'\|_\infty^2 + (1 + \pi)\|f'\|_\infty .
\end{equation}
The following corollary is an immediate consequence of the
previous Theorem and Corollary \ref{ascor} and shows the
asymptotical behavior of $\lambda_0$ for weak magnetic fields.

\begin{corollary} \label{c5.2}
If we replace the magnetic field $B$ by $\alpha B$, where $\alpha
\in \real$, then
\begin{equation}
\lambda_0 \geq \frac{\alpha^2}{2 k_4 k_9c_5(1 + d^2)} + \Ordo
(\alpha^4),
\end{equation}
as $\alpha \to 0$, where the constants are given in {\rm
(\ref{k_4})}, {\rm (\ref{c_5})} and {\rm (\ref{dandc_9})}.
\end{corollary}

Without loss of generality we assume that $\Omega_\lambda$
includes a small triangle spanned by the points $(-s,1),\, (s,1)$
and $(0,\pi(1 + \beta \lambda))$ with $s,\, \beta > 0$.

\begin{theorem} \label{t5.3}
Let the magnetic field $B$ be replaced by $\alpha B$, where
$\alpha \in \real$ and assume that
\begin{equation}
\alpha^2 \leq \frac{\pi s \beta}{4 \|A\|^2} \, \lambda + \Ordo
(\lambda^2),
\end{equation}
as $\lambda \to 0$, where $A$ is any magnetic vector potential
associated with $B$. Then the operator $M_d$ has at least one
eigenvalue below the essential spectrum.
\end{theorem}

\begin{proof}
Define the trial function $\varphi$ introduced in \cite{BGRS}, as
follows
\begin{equation}
\varphi (x,y) = \left\{
\begin{array}{l@{\quad , \ }l}
 \sin y\, e^{- s\beta\lambda (|x| - s)} & |x| \geq s,\ 0 < y < \pi, \\
 \sin \left( \frac y {1 + \beta\lambda \left( 1 - \frac {|x|} s \right)
  } \right) & |x| < s,\ 0 < y < \pi\left(1 + \beta\lambda \left( 1 - \frac
  {|x|} s\right)\right), \\
0 & \text{otherwise.}
\end{array}
\right.
\end{equation}
Let $\|\cdot\| = \|\cdot\|_{\Ltwo (\Omega_\lambda)}$. A simple
calculation gives
\begin{equation} \label{gradphi}
\frac{\|\nabla\varphi\|^2} {\|\varphi\|^2} \, = 1 - \lambda^2 \
\frac{s^2 \beta^2}{2}\, + \Ordo (\lambda^3),
\end{equation}
for $\lambda \to 0$. In order to prove that the discrete spectrum
of $M_d$ is non-empty, it is enough to show that the inequality
\begin{equation} \label{basic}
\frac{\|(\imunit\nabla+\alpha A)\varphi\|^2}{\|\varphi\|^2}\, < 1
\end{equation}
is satisfied for certain values of $\lambda$ and $\alpha$. By
(\ref{a_1}) and (\ref{a_2}) it follows that $|A| \in \Ltwo
(\Omega_\lambda)$. Since $\|\varphi\|_\infty = 1$, we have
\begin{equation}
\frac{\|(\imunit\nabla+\alpha A)\varphi\|^2}{\|\varphi\|^2}\, \leq
\frac{\|\nabla\varphi\|^2}{\|\varphi\|^2} +\frac{\alpha^2\,
\|A\|^2}{\|\varphi\|^2} = 1 - \lambda^2 \ \frac{s^2 \beta^2}{2}\,
+ \frac{\alpha^2\, \|A\|^2}{\|\varphi\|^2} + \Ordo (\lambda^3),
\end{equation}
Taking into account the fact that
\begin{equation}
\|\varphi\|^2 = \pi\left(\frac 1 {2 s \beta \lambda} + s +
\frac{\beta \lambda s}{2} \right)
\end{equation}
we get
\begin{equation}
\alpha^2 \leq \frac{\pi s \beta}{4 \|A\|^2} \, \lambda + \Ordo
(\lambda^2)
\end{equation}
and the proof is complete.
\end{proof}

We remark that Corollary \ref{c5.2} together with Theorem \ref{t5.3} show that
the order in the asymptotical behavior of the constant $c_H$ given
in Corollary \ref{ascor} is sharp.

\section{Locally curved waveguides} \label{curvedsection}

Let $a$ and $b$ be real-valued functions in $C^2(\real)$. Define
the set
\begin{equation} \label{thestarter}
\Omega_\gamma = \left\{ (s,t) \suchthat s = a(x) - y b'(x),\, t =
b(x) + y a'(x), \text{ where } (x,y) \in \real \times (0, \pi)
\right\},
\end{equation}
where $\gamma$ is to be explained later. We assume that
\begin{equation}
a'(x)^2 + b'(x)^2 = 1,
\end{equation}
for all $x \in \real$. The boundary of $\Omega_\gamma$ for which $y = 0$ is a
curve $\Gamma \in \real^2$ given by
\begin{equation}
\Gamma = \left\{(a(x),b(x)) \suchthat x \in \real \right\},
\end{equation}
and the signed curvature $\gamma : \real \to \real$ of $\Gamma$ is
given by
\begin{equation}
\gamma (x) = b'(x) a''(x) - a'(x) b''(x).
\end{equation}
Assume that $\gamma \in C_0^1(\real)$ and let the natural condition 
\begin{equation} \label{thefinnish}
\gamma(x) > - \frac 1{\pi},
\end{equation}
hold for all $x \in \real$. We prohibit $\Omega_\gamma$ to be
self-intersecting.

We will formulate the theory and results in terms of the curvature
$\gamma$ and not in terms of the functions $a$ and $b$. Those
functions $a$ and $b$ can be constructed from $\gamma$ uniquely up to
rotations and translations from the identities
\begin{eqnarray}
a(x) &=& a(0) + \int_{0}^{x} \cos \left( \int_{0}^{x_1} \gamma(x_2) \D
x_2 \right) \D x_1, \\
b(x) &=& b(0) + \int_{0}^{x} \sin \left( \int_{0}^{x_1} \gamma(x_2) \D
x_2 \right) \D x_1.
\end{eqnarray}

In $1994$, Duclos and Exner \cite{DE} gave a proof based on ideas
from Goldstone and Jaffe \cite{GJ} of existence of bound states
below the essential spectrum for the Schr\"odinger operator
$-\Delta$ in $\Omega_\gamma$ with Dirichlet boundary conditions,
assuming that $\gamma \neq 0$. Our aim is to prove that if we
introduce an appropriate magnetic field into the system it will
make the threshold of the bottom of the essential spectrum stable
if the curvature $\gamma$ is weak enough.

To be able to study weak curvatures we replace $\gamma$ by $\beta
\gamma$, where $\beta$ is a small positive real number. We will
use the notation $\Omega_\beta$ for the set $\Omega_{\beta
\gamma}$. Let $B \in C_0^1(\real^2)$ be a magnetic field such that
$B$ is not identically zero in $\Omega_\beta$. Let the operator
$M_c$ be the Friedrich's extension of the symmetric, semi-bounded
operator
\begin{equation}
\left( - \imunit \partial_s + a_1 \right)^2 + \left(- \imunit
\partial_t + a_2 \right)^2
\end{equation}
on the domain $\testfunc (\Omega_\beta)$, where $A(s, t) =
(a_1(s,t), a_2(s,t))$ is a magnetic vector potential associated
with $B$. Without loss of generality we can assume that $A$ is
defined by the identities (\ref{a_1}) and (\ref{a_2}). By
(\ref{Aisdecreasing}) and Theorem 
\ref{ess} we have $\essspectrum (M_c) = [1,\infty)$.

\begin{theorem} \label{curved}
There exists positive number $\beta_0$ depending on
$\|\gamma\|_\infty$, $\|\gamma'\|_\infty$, $\|a_1\|_\infty$ and
$\|a_2\|_\infty$ such that for $\beta \in (0, \beta_0)$ the
discrete spectrum of $M_c$ is empty.
\end{theorem}

\begin{proof}
The quadratic form $\form_c$ associated with $M_c$ is given by
\begin{equation}
\form_c [\psi] = \int_{\Omega_\beta} \left( |- \imunit \psi_s +
a_1 \psi|^2 + |- \imunit \psi_t + a_2 \psi|^2 \right) \D s \D t,
\end{equation}
on $\domain (\form_c) = H_0^1 (\Omega_\beta)$.
Define the unitary operator
\begin{equation} \label{U1}
U_\beta : \Ltwo (\Omega_\beta) \to \Ltwo (\Omega_0)
\end{equation}
as
\begin{equation} \label{U2}
\left(U_\beta \psi\right)(x,y) = \sqrt{1 + y \beta \gamma(x)} \
\psi(a(x) - y b'(x), b(x) + y a'(x)).
\end{equation}
The operator $M_c$ is unitary equivalent to the operator
\begin{equation}
M_\beta \bydef U_\beta M_c U_\beta^{-1}
\end{equation}
acting on the dense subspace $\domain (M_\beta) = U_\beta \domain
(M_c)$ of the Hilbert space $\Ltwo (\Omega_0)$. Our aim is to
prove that the operator $M_\beta - 1$ is nonnegative. For this we
calculate the quadratic form $\form_\beta$ associated with
$M_\beta$. Our change of variables gives us the Jacobian,
\begin{equation}
\frac {\partial \left(s, t\right)}{\partial \left(x, y\right)} =
\left( \begin{array}{cc} {a'(x) - y b''(x)} & {b'(x) + y a''(x)}
\cr {- b'(x)} & {a'(x)} \end{array} \right).
\end{equation}
Hence we have
\begin{equation}
\left\{
\begin{array}{lll}
\pdo_s &=& (1 + y \beta \gamma)^{-1} \left(a'(x)\pdo_x - (b'(x) +
y a''(x)) \pdo_y \right) \\ \pdo_t &=& (1 + y \beta \gamma)^{-1}
\left( b'(x) \pdo_x - (a'(x) - y b''(x)) \pdo_y \right)
\end{array}
\right.
\end{equation}
thus
\begin{align} \label{617}
\form_\beta [\varphi] &= \form_c [U_\beta^{-1} \varphi] \\
&= \int_{\Omega_0} \left( \left|\left[\frac { - \imunit \left( a'(x)
\pdo_x - (b'(x) + y a''(x)) \pdo_y \right)}{1 + y \beta \gamma(x)}
+ \tilde a_1(x,y) \right] \left( \frac{\varphi(x,y)} {\sqrt{1 + y
\beta \gamma(x)}} \right) \right|^2 \right. \nonumber  \\
& \left. \quad + \left| \left[ \frac {-\imunit\left( b'(x) \pdo_x
+ (a'(x) - y b''(x)) \pdo_y\right)}{1 + y \beta \gamma(x)} +
\tilde a_2(x,y) \right] \left( \frac{\varphi(x,y)}{\sqrt{1 + y
\beta \gamma(x)}} \right)\right|^2 \right) \nonumber \\
& \quad \quad (1 + y \beta \gamma(x)) \D x \D y, \nonumber
\end{align}
where
\begin{eqnarray} \label{ABfield2}
\tilde A(x,y) = (\tilde a_1(x,y), \tilde a_2(x,y)) = A(a(x) - y b'(x), b(x) +
y a'(x)).
\end{eqnarray}
We continue without writing arguments of the functions and use the
identities $a' a'' + b' b'' = 0$ and $(a'')^2 + (b'')^2 = \beta^2
\gamma^2$,
\begin{eqnarray} \label{qgamma}
\form_\beta [\varphi] &=& \int_{\Omega_0} \left( \frac
{|\varphi_x|^2}{(1 + y \beta \gamma)^2} - \frac{\imunit (a' \tilde
a_1 + b' \tilde a_2 )}{1 + y \beta \gamma} (\varphi_x \overline
\varphi - \varphi \overline{\varphi_x} ) + |\varphi_y|^2 \nonumber
\right.
\\
&& \qquad - \frac{\imunit \left( - (b' + y a'') \tilde a_1 + (a' - y b'')
    \tilde a_2\right)}{1 + y \beta \gamma} (\varphi_y \overline{\varphi}
    - \varphi \overline {\varphi_y} ) \\
&& \qquad - \frac{y \beta \gamma' }{2(1 + y \beta \gamma)^3}(\varphi
\overline{\varphi_x} + \varphi_x \overline \varphi) - \frac{\beta
\gamma}{2(1 + y \beta \gamma)} (\varphi \overline{\varphi_y} +
\varphi_y \overline{\varphi}) \nonumber \\
&& \left. \qquad + \left( \frac{y^2 \beta^2 \left(\gamma' \right)^2}
{4(1 + y \beta \gamma)^4} + \frac{\beta^2 \gamma^2}{4(1 + y \beta
\gamma)^2} + \tilde a_1^2 + \tilde a_2^2\right) |\varphi|^2\right)
\D x \D y. \nonumber
\end{eqnarray}

We write the form $q_\beta$ as a perturbation of the form
\begin{equation}
\form [\varphi] \bydef \int_{\Omega_0} |-\imunit \varphi_x + (a'
  \tilde a_1 + b' \tilde a_2) \varphi|^2 + |-\imunit \varphi_y +
  (- b' \tilde a_1 + a' \tilde a_2) \varphi|^2 \D x \D y,
\end{equation}
\ie
\begin{eqnarray}
\form_\beta [\varphi] - \|\varphi\|_{\Ltwo (\Omega_0)} ^2 &=&
    \form [\varphi] - \|\varphi\|_{\Ltwo (\Omega_0)} ^2
\label{curitfollow} \\
&& - \int_{\Omega_0} \bigg( \frac{2 y \beta \gamma + y^2 \beta^2
\gamma^2}{1 + y \beta \gamma} |\varphi_x|^2 - \imunit y \beta
\gamma (a' \tilde a_1 + b' \tilde a_2) (\varphi_x \overline\varphi
- \varphi\overline{\varphi_x}) \nonumber \\
&& \qquad - \imunit y \left(- \beta \gamma b' \tilde a_1 + \beta
\gamma a' \tilde a_2 + \frac{a'' \tilde a_1 - b'' \tilde a_2} {1 +
y \beta \gamma}\right) \left( \varphi_y \overline \varphi -
\varphi \overline {\varphi_y} \right) \nonumber \\
&& \qquad + \frac{y \beta \gamma'}{2(1 + y \beta \gamma)^3}
\left(\varphi\overline{\varphi_x} + \varphi_x \overline \varphi
\right) + \frac{\beta \gamma}{2(1 + y \beta \gamma)} \left(\varphi
\overline{\varphi_y} + \varphi_y \overline \varphi \right)
\nonumber \\
&& \qquad - \left( \frac{y^2 \beta^2 \left( \gamma' \right)^2}
{4(1 + y \beta \gamma)^4} + \frac{\beta^2 \gamma^2} {4(1 + y \beta
\gamma)^2} \right) |\varphi|^2 \bigg) \D x \D y. \nonumber
\end{eqnarray}
We can easily arrive at the following estimate
\begin{eqnarray}
\form_\beta [\varphi] - \|\varphi\|_{\Ltwo (\Omega_0)} ^2
&\geq& \form [\varphi] - \|\varphi\|_{\Ltwo (\Omega_0)} ^2 \\
&& - \beta \int_{\Omega_0} \chi \left( c_{10} \left(|\varphi_x|^2 +
|\varphi_y|^2 \right) + c_{11} |\varphi|^2 \right) \D x \D y,
\nonumber
\end{eqnarray}
where $\chi$ is the characteristic function of the support of
$\gamma$ and
\begin{eqnarray}
c_{10} &=& \pi^2 \|\gamma\|_\infty^2 + 2 \pi \left(1 +
\|a_1\|_\infty + \|a_2\|_\infty \right) \|\gamma\|_\infty +
\frac{\pi}{2} \|\gamma'\|_\infty, \\ 
c_{11} &=& \left(\frac 12 + 3 \pi \|a_1\|_\infty + 3
\pi \|a_2\|_\infty\right) \|\gamma\|_\infty + 
\frac \pi 2 \|\gamma'\|_\infty.
\end{eqnarray}
By using (\ref{generalhardyinequality}) and (\ref{abtoa2b2}) we
get
\begin{eqnarray}
\form_\beta [\varphi] - \|\varphi\|_{\Ltwo (\Omega_0)}^2 &\geq&
\left( \frac 12 - 2 \beta c_{10} \right) \left( \form [\varphi] -
\|\varphi\|_{\Ltwo (\Omega)} ^2 \right) \\
&& + \left( \frac {c_H}{2} - \beta c_{12} (1 + d^2)\right) \int_{\Omega_0}
    \frac{|\varphi|^2}{1 + x^2} \D x \D y, \nonumber
\end{eqnarray}
where
\begin{equation} \label{dandc_12}
d = \max \supp \gamma \quad \text{ and }
\quad c_{12} = 2 \left(1 + \|a_1\|_\infty^2 + \|a_2\|_\infty^2
\right) c_{10} + c_{11}.
\end{equation}
The right hand side is positive if $\beta \in (0, \beta_0)$, with
\begin{equation}
\beta_0 \bydef \frac{c_H}{2c_{12}(1 + d^2)}.
\end{equation}
Hence the operator $M_d$ has empty discrete spectrum.
\end{proof}

If we replace $B$ by $\alpha B$, $A$ will be replaced $\alpha A$. Let us
define
\begin{equation}
k_{12} := \lim_{\alpha\to 0} c_{12} = 2 \pi^2 \|\gamma\|_\infty^2 +
  (4 \pi + 2^{-1}) \|\gamma\|_\infty + \frac{3\pi}{2}\|\gamma'\|_\infty.
\end{equation}

\begin{corollary}
If we replace the magnetic field $B$ by $\alpha B$, where $\alpha
\in \real$, then
\begin{equation}
\beta_0 \geq \frac{\alpha^2}{2 k_4 c_5 c_{12}(1 + d^2)} + \Ordo
(\alpha^4),
\end{equation}
as $\alpha \to 0$, where the constants are given in {\rm
(\ref{k_4})}, {\rm (\ref{c_5})} and {\rm (\ref{dandc_12})}.
\end{corollary}

\section{Aharonov-Bohm field}

In this last section we consider the Aharonov-Bohm field. The field is
generated by a magnetic vector potential having a singularity in one point.

\subsection{A Hardy-type inequality}

Let $p$ be the point $(0,y_0) \in \real^2$, where $y_0 \in
(0,\pi)$ and define $A : \real^2 \to \real^2$ to be the vector
field
\begin{equation} \label{ABfield}
A(x,y) = (a_1(x,y), a_2(x,y)) = \Phi \cdot \left( \frac {- y +
  y_0}{x^2 + (y - y_0)^2}, \frac x{x^2 + (y - y_0)^2}\right),
\end{equation}
for $\Phi \in \real$. The vector field $A$ is a magnetic vector
potential for the Aharonov-Bohm magnetic field. The magnetic field
$B : \real^2 \to \real$ is for $(x,y) \neq p$ given by
\begin{equation}
B(x,y) = \pdo_x a_2 - \pdo_y a_1 = 0
\end{equation}
and the constant $2\pi\Phi$ is the magnetic flux through the point $p$,
\ie let $\Gamma$ be a closed simple curve containing $p$, then
\begin{equation}
\oint_\Gamma a_1 \D x + a_2 \D y = 2 \pi \Phi.
\end{equation}
Let $\Omega \subset \real^2$ be given by $\Omega = \real \times
(0,\pi)$. The following Hardy-inequality holds true.

\begin{theorem} \label{hardytheorem}
Let $A \in L^2_{\loc} (\real^2)$ be a given real-valued
magnetic vector potential such that there exists a ball
$\ball_R(p) \subset \Omega$, for which $(x,y) \in \ball _R(p)$
implies that
\begin{equation}
A (x,y) = \Phi \cdot \left(\frac {-y + y_0}{x^2 + (y
    - y_0)^2}, \frac x{x^2 + (y - y_0)^2} \right),
\end{equation}
where $\Phi \in \real \setminus \zahl$. Then for all $v \in
H_{0,A}^1 (\Omega \setminus \{p\})$ the following inequality holds
\begin{eqnarray} \label{hardyinequality}
c_{AB} \int_{\Omega} \frac{|v|^2 \D x \D y}{x^2 + (y - y_0)^2}
&\leq& \int_{\Omega} \left( |-\imunit \nabla v + Av|^2 - |v|^2 \right) \D x \D
  y,
\end{eqnarray}
where
\begin{align} \label{hardyconstant}
c_{AB} &= \frac{R^2 \Psi^2 \cos^2 \left(\left|y_0 - \frac \pi 2\right| + R
  \right)} {8 \left(2 R^2 \Psi^2 + (2c_{13} \Psi^2 + 1 + 2c_{13})(9 R^2 + 16
\pi^2)\right)},
\\
\Psi &= \min_{k\in\zahl} |\Phi - k|, \label{psiconstant}\\
c_{13} &= \frac{4 \pi^2}{\pi^2 - \max \left\{ y_0^2, (\pi -y_0)^2 \right\}}. 
\label{c_13}
\end{align}
\end{theorem}

For the proof of the Theorem we need two lemmas.

\begin{lemma} \label{circleestimatelemma}
Let $R$ be chosen such that $\ball_R(p) \subset \Omega$, then the
inequality
\begin{equation} \label{circleestimate}
\int_{\ball_R(p)} |-\imunit\nabla u + A u|^2 \sin^2 y \D x \D y
\geq \Psi^2 \int_{\ball_R(p)} \frac{\cos^2 \left(|(x,y) - p| \right)
|u|^2 \sin^2 y \D x \D y} {x^2 + (y - y_0)^2}
\end{equation}
holds true for all $u \in C^\infty(\overline{\ball_R(p)})$ such that $u = 0$
in a neighborhood of $p$, where $\Psi$ is given in {\rm (\ref{psiconstant})}.
\end{lemma}

\begin{proof}
We follow ideas from \cite{LW}. Let us introduce polar coordinates
centered at the point $p$ and let $D_n = \left\{ (r,\theta)
\suchthat (n-1)RN^{-1}< r
  < nRN^{-1}\right\}$, where $N$ is a natural number. Let $u \in C^\infty
(\overline {\ball_R(p)})$ such that $u = 0$ in a neighborhood of $p$. In each
$D_n$ we have 
\begin{align}
\int_{D_n} |-\imunit \nabla u + A u|^2 \sin^2 y \D x \D y &=
\int_{D_n} \left( |u_r|^2 + r^{-2}|-\imunit u_\theta + \Phi u|^2
\cos^2 (r \sin
  \theta) \right)r \D r \D \theta \nonumber \\
&\geq \cos^2 \left(\frac {nR}{N} \right) \int_{D_n} r^{-1}|-\imunit u_\theta
 + \Phi u|^2 \D r \D \theta. \label{TiALform}
\end{align}

To study the form (\ref{TiALform}) we make use of the
one-dimensional self-adjoint operator $K$ on $\Ltwo (0,\pi)$ given
by
\begin{equation}
K = -\imunit \pdo_{\theta} + \Phi,
\end{equation}
defined on the set
\begin{equation}
\domain (K) = \left\{u \in H^1(0, 2\pi) \suchthat u(0) =
u(2\pi)\right\}.
\end{equation}
The spectrum of $K$ is discrete and its eigenvalues
$\left\{\lambda_k\right\}_{k \in \zahl}$ and the complete
orthonormal system of eigenfunctions $\left\{\varphi_k\right\}_{k
\in \zahl}$ are given by
\begin{equation}
\lambda_k = k + \Phi
\end{equation}
and
\begin{equation}
\varphi_k(\theta) = \frac 1{\sqrt{2\pi}}e^{\imunit \theta
(\lambda_k - \Phi)}.
\end{equation}
We can write the function $u$ in the Fourier expansion
\begin{equation}
u(r,\theta) = \sum_{k \in \zahl} \omega_k(r) \varphi_k(\theta).
\end{equation}
Then we have
\begin{eqnarray}
\int_{D_n} r^{-1} |-\imunit u_\theta + \Phi u|^2 \D r \D \theta
&\geq& \int_{D_n} r^{-1} \left| \sum_{k \in \zahl} \omega_k \lambda_k
  \varphi_k \right|^2 \D \theta \D r \nonumber \\
&\geq& \int_{(n-1)RN^{-1}}^{nRN^{-1}} r^{-1} \sum_{k \in \zahl} |\omega_k|^2
  \lambda_k^2 \D r \nonumber \\
%&\geq& \Psi^2 \cos^2 R \int_0^R r^{-1} \sum_{k \in \zahl}
%  |\omega_k|^2 \, \D r \\
&\geq& \Psi^2 \int_{D_n} r^{-1} |u|^2 \D r \D \theta.
%&\geq& \Psi^2 \cos^2 R \int_{\ball_R(p)} \frac{|u|^2 \sin^2 y \D x \D y}
%{x^2+(y-\frac{\pi}2)^2} \nonumber \\
%&\geq \Psi^2 \int_{D_n} \frac{|u|^2 \sin^2 y \D x \D y}
%  {x^2 + (y - \frac{\pi}2)^2}. \nonumber
\end{eqnarray}
Finally we sum up the inequality over the rings. For any $N$ we
have
\begin{eqnarray}
\int_{\ball_R(p)} |-\imunit\nabla u + A u|^2 \sin^2 y \D x \D y
&\geq& \sum_{n=1}^N \cos^2 \left( \frac{nR}N \right) \int_{D_n}
r^{-1} | - \imunit u_\theta + \Phi u |^2 \D r \D \theta \nonumber
\\
&\geq& \sum_{n=1}^N \cos^2 \left( \frac{nR}N \right) \Psi^2 \int_{D_n} r^{-1}
|u|^2 \D r \D \theta \\
&\geq& \Psi^2 \sum_{n=1}^N \int_{D_n} \cos^2 \left(r + \frac{R}{N}
\right)r^{-1} |u|^2 \D r \D \theta. \nonumber
\end{eqnarray}
Hence the desired result will follow as $N \to \infty$.
\end{proof}

\begin{lemma} \label{onedimlemma}
The inequality
\begin{equation}
\int_0^{\pi} \frac{|u(y)|^2 \sin^2 y \D y}{(y - y_0)^2}
\leq c_{13} \int_0^{\pi} |u'(y)|^2 \sin^2 y \D y
\end{equation}
holds true for all functions $u \in H^1(0,\pi)$ such that
$u\left(\frac {\pi}2\right) = 0$, where $c_{13}$ is given by {\rm
  (\ref{c_13})}. 
\end{lemma}
\begin{proof}
It is clear that
\begin{equation}\label{operator15}
\pi^2 \min \left\{ y_0^{-2}, (\pi - y_0)^{-2}\right\} \ \leq \ -
\frac{d^2}{dy^2}.
\end{equation}
We will prove another estimate for $-\frac{d^2}{dy^2}$, namely the
inequality
\begin{equation} \label{317}
\frac{1}{4(y - y_0)^2} \ \leq \ -\frac{d^2}{dy^2},
\end{equation}
for the subspace of functions $v \in \testfunc (0,\pi)$ satisfying
$v(y_0) = 0$. It will be enough to prove that
\begin{equation}\label{anotherinequality}
\frac 14 \int_0^\beta \frac{|v(y)|^2 \D y}{y^2} \leq \int_0^\beta
|v'(y)|^2 \D y,
\end{equation}
for all functions $v \in \testfunc (0, \beta)$, where $\beta$ is
any positive number.

Let $v \in \testfunc (0,\beta)$ be a real-valued function, then
\begin{equation}
|v(y)|^2 = 2 \int_0^y v(t) v'(t) dt.
\end{equation}
Hence
\begin{eqnarray}
\int_0^\beta \frac{|v(y)|^2 \D y}{y^2} %&=& 2 \int_0^\beta \int_0^y
%\frac{v(t)v'(t) \D t \D y}{y^2} \nonumber \\
%&=& 2 \int_0^\beta \int_t^\beta \frac{v(t)v'(t) \D y \D t}{y^2} \nonumber
%\\
&=& 2 \int_0^\beta v(t)v'(t) \left( \frac 1t-\frac 1\beta\right) \D t \\
&\leq& 2 \left( \int_0^\beta |v(t)|^2 \left( \frac 1t - \frac 1\beta \right)^2
  \D t \right)^{\frac 12} \left( \int_0^\beta |v'(t)|^2 \D t
\right)^{\frac 12} \nonumber \\
&\leq& 2 \left( \int_0^\beta \frac{|v(t)|^2 \D t}{t^2} \right)^{\frac 12}
\left( \int_0^\beta |v'(t)|^2 \D t \right)^{\frac 12} \nonumber
\end{eqnarray}
from what (\ref{anotherinequality}) follows. The estimates
(\ref{operator15}) and (\ref{317}) imply that
\begin{equation}
\frac{1} {(y - y_0)^2} \ \leq \ c_{13} \left(- \frac{d^2}{dy^2} - 1\right),
\end{equation}
which in terms of the quadratic form means that
\begin{equation}\label{sinsub115}
\int_0^{\pi} \frac{|v(y)|^2 \D y}{(y - y_0)^2} \ \leq \ c_{13}
\int_0^{\pi} |v'(y)|^2-|v(y)|^2 \D y,
\end{equation}
holds for all $v \in H_0^1 (0,\pi)$ such that $v(y_0) = 0$. The
substitution $v(y) = u(y) \sin y$ implies that $u \in H^1(0,\pi)$
and that $u(y_0) = 0$. From (\ref{sinsub115}) we get
\begin{equation}\label{onedimineq}
\int_0^{\pi} \frac{|u(y)|^2 \sin^2 y \D y}{(y -
y_0)^2} \ \leq \ c_{13} \int_0^{\pi} |u'(y)|^2 \sin^2 y \D y,
\end{equation}
for functions $u \in H^1 (0,\pi)$ such that $u(y_0) = 0$.
\end{proof}

Now we are in position to prove Theorem \ref{hardytheorem}. Since
the method used in the proof doesn't give a sharp constant we will
not put an effort in using optimal inequalities with the risk of
being lost in technicalities.

\begin{proof}[Proof of Theorem \ref{hardytheorem}]
If we substitute $v(x,y) = u(x,y) \sin y$ then inequality
(\ref{hardyinequality}) becomes
\begin{equation} \label{newwin}
c_{AB} \int_{\Omega} \frac{|u|^2 \sin^2 y \D x \D y}{x^2 + (y -
y_0)^2} \ \leq \
  \int_{\Omega} |-\imunit \nabla u + Au|^2 \sin^2 y \D x \D y.
\end{equation}
We need to prove the inequality (\ref{newwin}) for all $u \in
C^\infty (\overline \Omega) \cap \Ltwo (\Omega)$ such that $u = 0$ in a
neighborhood of the point $p$.

Define for $R \in (0, \dist (y_0, \partial\Omega))$ the set
$\Omega_R = (-R,R) \times (0,\pi)$ and let $h_\pm (x) = y_0 \pm
\sqrt{R^2 - x^2}$. Assume $x \in (-R,R)$, $x \neq 0$, $\psi$ is defined by
(\ref{smallpsi}) and let $u \in C^\infty (\overline{\Omega}) \cap \Ltwo
(\Omega)$ such that $u = 0$ in a neighborhood of the point $p$. Since $u(x,
\cdot) \psi \in H^1(0,\pi)$ we have by Lemma \ref{onedimlemma} that
\begin{eqnarray}
\int_0^{\pi} \frac{|u|^2 \sin^2 y \D y} {x^2 + (y - y_0)^2}
&\leq& 2c_{13} \int_0^{\pi} \left|u_y \psi + u \psi' \right|^2 \sin^2 y \D y +
2 \int_{h_-(x)}^{h_+(x)} \, \frac{|u|^2 \sin^2 y \D y}{x^2 + (y - y_0)^2}
\nonumber \\
&\leq& 4c_{13} \int_0^{\pi} |u_y|^2 \sin^2 y \D y \\
&& + \left( 2 + \frac{4c_{13}R^2}{R^2-x^2} \right) \int_{h_-(x)}^{h_+(x)}
\frac{|u|^2 \sin^2 y \D y}{x^2+(y - y_0)^2}, \nonumber
\end{eqnarray}
where $c_{13}$ is given by (\ref{c_13}). Thus the inequality
\begin{eqnarray}
\int_0^{\pi} \frac{|u|^2 (R^2 - x^2) \sin^2 y \D y} {x^2 + (y - y_0)^2}
&\leq& 4 c_{13} R^2 \int_0^{\pi} |u_y|^2 \sin^2 y \D y \label{229} \\
&& + \, 2R^2(1 + 2c_{13}) \int_{h_-(x)}^{h_+(x)} \frac{|u|^2 \sin^2 y \D
  y}{x^2 + (y - y_0)^2}, \nonumber
\end{eqnarray}
holds. By continuity the inequality can be extended to $u(x,\cdot) \in
H^1(0,\pi)$. We will make use of the diamagnetic inequality
(\ref{diamagneticinequality}), for functions $v \in H^1_{0,A} 
(\Omega \setminus \{p\})$. Let $u(x, \cdot) = |w(x, \cdot)|$, where $w \in
C^\infty (\overline{\Omega}) \cap \Ltwo (\Omega)$ such that $w = 0$ in a
neighborhood of the point $p$, then $u(x, \cdot) \in H^1(0,\pi)$ and by
Lemma \ref{circleestimatelemma}, (\ref{diamagneticinequality}) and (\ref{229})
we have
%\begin{eqnarray}
%\int_0^{\pi} \frac{|w|^2 (R^2-x^2) \sin^2 y \D y} {x^2 + (y - y_0)^2}
%\label{integratingfromRR}
%&\leq& 4c_{13}R^2 \int_0^{\pi} |-\imunit\nabla w + Aw|^2 \sin^2 y \D y \\
%&& + \, 2R^2(1 + 2c_{13}) \int_{h_-(x)}^{h_+(x)} \frac{|w|^2 \sin^2 y \D
%  y} {x^2 + (y - y_0)^2} \nonumber .
%\end{eqnarray}
%Integration the inequality (\ref{integratingfromRR}) from $-R$ to $R$
%w.r.t. $x$ and using Lemma \ref{circleestimatelemma} gives us
\begin{eqnarray}
\int_{\Omega_R} \frac{|w|^2 (R^2-x^2) \sin^2 y \D x \D y} {x^2 + (y - y_0)^2}
&\leq& \label{equivalentinnorm}
c_{14} \int_{\Omega_R} |-\imunit\nabla w + Aw|^2 \sin^2 y \D x \D y,
\end{eqnarray}
where the constant 
\begin{equation}
c_{14} = \frac{ 4 R^2 \Psi^2 c_{13} + 2R^2 + 4R^2c_{13}} {\Psi^2 \cos^2
  (|y_0 - \frac \pi 2| + R)}.
\end{equation}
Let $m = \frac R{\sqrt 2}$ and define $\varphi$ by (\ref{smallphi}). For $u
\in C^\infty (\overline \Omega) \cap \Ltwo (\Omega)$ such that $u$
vanishes in a neighborhood of the point $p$, $y \in (0,\pi)$, $y \neq y_0$,
we write $u = u \varphi + u (1 - \varphi)$ and use (\ref{classicalhardy}) to
get
\begin{eqnarray}
 \int_{-\infty}^{\infty} \frac{|u|^2 \D x}{x^2 + (y - y_0)^2} &\leq& 
16 \int_{-\infty}^{\infty} |u_x|^2 \D x + 16 \int_{-m}^m |u \psi
   '|^2 \D x \nonumber \\
&& + \, 2 \int_{-m}^m \frac{|u|^2 \D x}{x^2+\left(y-\frac{\pi}2 \right)^2} \\
&=& 16 \int_{-\infty}^{\infty} |u_x|^2 \D x + c_{15} \int_{-m}^m \frac{|u|^2
   \D x}{x^2 + (y - y_0)^2}, \nonumber
\end{eqnarray}
where $c_{15} = 18 + \frac {32\pi^2} {R^2}$. Since $y \neq y_0$ the inequality
can by continuity be extended to functions $u(\cdot,y) \in H_0^1 (\real)$. By
using (\ref{diamagneticinequality}) one gets
\begin{eqnarray}
\int_{-\infty}^{\infty} \frac{|u|^2 \D x}{x^2 + (y - y_0)^2} &\leq& 16
  \int_{-\infty}^{\infty} |-\imunit \nabla u + Au|^2 \D x \label{now248} \\
&& + \, c_{15} \int_{-m}^m \frac{ |u|^2 \D x}{x^2 + (y - y_0)^2}, \nonumber
\end{eqnarray}
for all $u \in C^\infty (\overline \Omega) \cap \Ltwo (\Omega)$ such that $u =
0$ in a neighborhood of $p$. Combining the inequalities
(\ref{equivalentinnorm}) and (\ref{now248}) we have 
\begin{eqnarray}
\int_{\Omega} \frac{|u|^2 \sin^2 y \D x \D y}{x^2 + (y - y_0)^2} &\leq& 
%16 \int_{\Omega} |-\imunit \nabla u + Au|^2 \sin^2 y \D x \D y \nonumber \\
%&& + \, c_3 \int_{\Omega_m} \frac{|u|^2 \sin^2 y \D x \D y} {x^2 + (y -
%  y_0)^2} \nonumber \\
%&\leq& 16 \int_{\Omega} |-\imunit \nabla u + Au|^2 \sin^2 y \, \D x \D y
%  \label{almdone} \\
%&& + \frac{c_3}{R^2 - m^2} \int_{\Omega_R} \frac{|u|^2(R^2 - x^2) \sin^2 y \D
%  x \D y}{x^2 + (y - y_0)^2} \nonumber \\
%&=& 
c_{16} \int_{\Omega} |-\imunit \nabla u + Au|^2 \sin^2 y \, \D x \D y,
\end{eqnarray}
where the constant $c_{16} =
16 + \frac{2c_{14}c_{15}}{R^2}$. This proves the inequality
(\ref{newwin}) with the constant $c_{AB} = c_{16}^{-1}$.
\end{proof}

\subsection{Locally deformed waveguides}

Let $f$ be a non-negative function in $C^1_0 (\real)$ and for $\lambda \geq 0$
we define
\begin{equation}
\Omega_\lambda = \left\{(x,y) \in \real^2 \suchthat 0 < y < \pi + \lambda \pi
  f(x)\right\} \setminus \{p\},
\end{equation}
where $p = (0,y_0)$. Let $M_d$ be the Friedrich's
extension of the symmetric, semi-bounded operator
\begin{equation}
\left( -\imunit \pdo_s + a_1(s,t)\right)^2 + \left( -\imunit
\pdo_t + a_2(s,t)\right)^2,
\end{equation}
on the domain $\testfunc (\Omega_\lambda)$, where the magnetic
vector potential is for $\Phi \in \real \setminus \zahl$ is defined by
(\ref{ABfield}). For simplicity we assume that $\supp f \subset [\frac \pi
2,\infty)$. Since $\divergence A = 0$ and $|A| \in \Ltwo ((1, \infty) \times
(0, \pi))$ we have by Theorem \ref{ess} that the essential spectrum of $M_d$ 
equals $[1,\infty)$. The following Theorem says that the spectrum of $M_d$ is
stable under small deformations. 

\begin{theorem}
There exists a value $\lambda_0$ depending on $\|f\|_\infty$ and
    $\|f'\|_\infty$ such that for $\lambda \in (0,\lambda_0)$ the discrete
    spectrum of $M_d$ is empty. 
\end{theorem}
\begin{proof}
Let the unitary mapping $U_\lambda$ be given by (\ref{u1}) and (\ref{u2}).
The operator $M_d$ is unitary equivalent to
\begin{equation}
M_\lambda \bydef U_{\lambda} M_d U^{-1}_{\lambda},
\end{equation}
defined on the set $U_{\lambda}\domain (M_d)$ in $\Ltwo (\Omega)$.
The quadratic form associated with $M_d$ is 
\begin{equation}
\form_d [\psi] = \int_{\Omega_\lambda} |-\imunit \psi_s + a_1
  \psi|^2 + |-\imunit \psi_t + a_2 \psi|^2 \D s \D t,
\end{equation}
defined on $\domain (\form_d) = H^1_{0,A}(\Omega_\lambda)$. Hence the form
associated with $M_\lambda$ is 
\begin{equation}
\form_\lambda [\varphi] = \form_d [U^{-1}_\lambda \varphi]
\end{equation}
defined on the space $\domain (\form_\lambda) = U_{\lambda} \domain (\form_d)$.

Since $\essspectrum (M_\lambda) = \essspectrum (M_d) = [1, \infty)$ it will be
enough to prove that $M_\lambda - 1$ is non-negative. 
Let $g(s) = 1 + \lambda f(s)$ and let $\form$ be the quadratic form associated
with the Schr\"odinger operator with the magnetic vector potential $\tilde A$
in the space $\Ltwo (\Omega_0)$. Without loss of generality we assume that
$\lambda \leq 1$. It follows from (\ref{generaldeform}) that
\begin{eqnarray} \label{417}
\form_\lambda [\varphi] - \|\varphi\|_{\Ltwo (\Omega_0)}^2
&=& \form [\varphi] - \|\varphi\|_{\Ltwo (\Omega_0)}^2
\nonumber \\
&& + \int_{\Omega_0} \left( \frac{y^2 \lambda^2 (f')^2 - 2 \lambda
  f - \lambda^2f^2}{g^2}|\varphi_y|^2 - \frac 14 \left(\frac{\lambda
    f'}{g} \right)^2|\varphi|^2 \right. \nonumber \\
&& \qquad - \frac{y\lambda f'}{g} (\varphi_x \overline{\varphi_y}
+ \varphi_y \overline{\varphi_x}) - \frac{\lambda f'}{2g}
(\varphi\overline{\varphi_x} +
   \varphi_x \overline{\varphi}) \nonumber \\
&& \qquad \left. + \imunit \lambda \frac{yf' \tilde a_1 + f \tilde a_2}g (
\varphi_y \overline{\varphi} - \varphi\overline{\varphi_y})
\right) \D x \D y \nonumber \\
&\geq& \form [\varphi] - \|\varphi\|_{\Ltwo (\Omega_0)}^2 \\
&& - \lambda \int_{\Omega_0} \chi \cdot \big( c_{17} \left( |\varphi_x|^2 +
|\varphi_y|^2 \right) + \left(c_{18} + c_{19}(\tilde a_1^2 + \tilde
  a_2^2)\right)|\varphi|^2\big) \D x \D y, \nonumber
\end{eqnarray}
where $c_{17} = 2\pi \|f'\|_\infty + 3 \|f\|_\infty +
\|f\|_\infty^2$, $c_{18} = \frac 14 \|f'\|_\infty^2 + \frac 12
\|f'\|_\infty$, $c_{19} = \pi \|f'\|_\infty + \|f\|_\infty$ and
$\chi$ is the characteristic function of the support of $f$. From
(\ref{abtoa2b2}) we get
\begin{eqnarray}
\form_\lambda [\varphi] - \|\varphi\|_{\Ltwo (\Omega_0)}^2
&\geq& \form [\varphi] - \|\varphi\|_{\Ltwo (\Omega_0)}^2 \nonumber \\
&& - \lambda \int_{\Omega_0} \chi \bigg( 2c_{17} \left(|-\imunit\nabla \varphi
  + \tilde A \varphi|^2 - |\varphi|^2 \right) \\
&& + \left( \frac {(2c_{17} + c_{18})(d^2 + \pi^2)}{x^2 + \left(y -
    y_0\right)^2} + (2c_{17} + c_{19})(\tilde a_1^2 + \tilde 
a_2^2) \right)|\varphi|^2 \bigg) \D x \D y, \nonumber
\end{eqnarray}
where $d = \max \supp f$. We use the pointwise inequality
\begin{equation}
\chi (x) \cdot \left( \tilde a_1^2(x,y) + \tilde a_2^2(x,y) \right)
  \leq \frac {4 \Phi^2 \left(d^2 + \pi^2\right)} {\pi^2(x^2 + \left(y
  - y_0)\right)^2}
\end{equation}
to get
\begin{eqnarray}
\form_\lambda [\varphi] - \|\varphi\|_{\Ltwo (\Omega_0)}^2
&\geq& \form [\varphi] - \|\varphi\|_{\Ltwo (\Omega_0)}^2 - 2 \lambda c_{17}
\left(\form [\varphi] - \|\varphi\|_{\Ltwo (\Omega_0)}^2 \right) \\
&& - \lambda \int_{\Omega_0} \frac {c_{20} d^2 + c_{21}}{x^2 + \left(y -
    y_0\right)^2} |\varphi| ^2 \D x \D y, \nonumber 
\end{eqnarray}
where $c_{20} = 2c_{17} + c_{18} + 4 \Phi^2 \pi^{-2} (2c_{17} + c_{19})$ and
$c_{21} = \pi^2 (2c_{17} + c_{18}) + 4 \Phi^2 (2c_{17} + c_{19})$. From
Theorem \ref{hardytheorem} we have 
\begin{eqnarray}
\form_\lambda [\varphi] - \|\varphi\|_{\Ltwo (\Omega_0)}^2
&\geq& \left( \frac 12 - 2 \lambda c_{17} \right) \left( \form
[\varphi] - \|\varphi\|_{\Ltwo (\Omega_0)}^2 \right) \nonumber \\
&& + \left( \frac {c_{AB}}2 - \lambda \left( c_{20} d^2 + c_{21} \right)
\right) \int_{\Omega_0} \frac {|\varphi|^2}{x^2 + \left( y - y_0\right)^2} \D 
x \D y \nonumber \\
&\geq& 0, \nonumber
\end{eqnarray}
for $\lambda \in (0, \lambda_0)$, where $c_{AB}$ is the constant from
(\ref{hardyconstant}) and
\begin{equation}
\lambda_0 = \frac{c_{AB}}{2(c_{20} d^2 + c_{21})}.
\end{equation}
\end{proof}

\subsection{Locally curved waveguides}

Let $A$ be given as in (\ref{ABfield}) and let $\Omega_\gamma$ be defined by
(\ref{thestarter}) -- (\ref{thefinnish}) with the additional assumption that
$a(x) = x$ and $b(x) = 0$ for $x \leq \frac \pi 2$. To be able to study weak
curvatures we replace $\gamma$ by $\beta \gamma$ for arbitrary $\beta \geq
0$. We denote by $\Omega_\beta$ the set $\Omega_{\beta \gamma}$. 
\begin{equation}
\form_c [\psi] \bydef \int_{\Omega_\beta} | - \imunit \psi_s +
a_1 \psi|^2 + | - \imunit \psi_t + a_2 \psi|^2 \D s \D t,
\end{equation}
be defined on $\domain (\form_c) = H_{0,A}^1(\Omega_\beta)$. Then
  $\form_c$ is the quadratic form associated with the Friedrich's
  extension $M_c$ of the the symmetric, semi-bounded operator
\begin{equation}
\left( - \imunit \pdo_s + a_1(s,t)\right)^2 + \left( -\imunit
\pdo_t + a_2(s,t)\right)^2,
\end{equation}
defined on $\testfunc (\Omega_\beta)$. For simplicity we assume that $\supp
\gamma \subset [\frac \pi 2 , \infty)$. By Theorem \ref{ess} we get that the
essential spectrum of $M_c$ equals $[1,\infty)$.

\begin{theorem}
There exists a positive number $\beta_0$ such that for $\beta \in (0,
 \beta_0)$ the discrete spectrum of $M_c$ is empty.
\end{theorem}

\begin{proof}
Denote by $M_\beta$ the operator $U_\beta M_c U^{-1}_\beta$,
where $U_\beta$ is defined in (\ref{U1}) and (\ref{U2}). Let
$\form_\beta$ be the form associated with $M_\beta$ defined on the domain
$\domain (\form_\beta) = U_\beta \domain (\form_c)$. Following the
calculations in (\ref{617}) -- (\ref{curitfollow}) we get
\begin{eqnarray} \label{newqe}
\form_\beta [\varphi] - \|\varphi\|_{\Ltwo (\Omega_0)} ^2 &=&
    \form [\varphi] - \|\varphi\|_{\Ltwo (\Omega_0)}^2 \\
&& - \int_{\Omega_0} \bigg( \frac{2 y \beta \gamma + y^2 \beta^2 \gamma^2}{1 +
    y \beta \gamma} |\varphi_x|^2 - \imunit y \beta \gamma (a' \tilde a_1 + b'
    \tilde a_2) (\varphi_x\overline \varphi - \varphi\overline{\varphi_x} )
    \nonumber \\ 
&& \qquad - \imunit y \left(- \beta \gamma b' \tilde a_1 + \beta \gamma a'
    \tilde a_2 + \frac{a'' \tilde a_1 - b'' \tilde a_2} {1 + y \beta
    \gamma}\right) \left( \varphi_y \overline \varphi - \varphi \overline
    {\varphi_y} \right) \nonumber \\ 
&& \qquad + \frac{y \beta \gamma'}{2(1 + y \beta \gamma)^3} \left( \varphi
\overline{\varphi_x} + \varphi_x \overline \varphi \right) + 
    \frac{\beta \gamma}{2(1 + y \beta \gamma)} \left( \varphi
    \overline{\varphi_y} + \varphi_y \overline \varphi \right)  \nonumber \\
&& \qquad - \left( \frac{y^2 \beta^2 \left(\gamma' \right)^2}{4(1 + y \beta
    \gamma)^4} + \frac{\beta ^2 \gamma ^2}{4(1 + y \beta \gamma)^2} \right)
    |\varphi|^2 \bigg) \D x \D y . \nonumber
\end{eqnarray}
Without loss of generality we can assume that $\beta \leq 1$, hence
\begin{eqnarray}
\form_\beta [\varphi] - \|\varphi\|_{\Ltwo (\Omega_0)} ^2
&\geq& \form [\varphi] - \|\varphi\|_{\Ltwo (\Omega_0)} ^2 \\
&& - \beta \int_{\Omega_0} \chi \left( c_{22} \left( |\varphi_x|^2 +
|\varphi_y|^2 \right) + \left(c_{23} + c_{24} (\tilde a_1^2 + \tilde a_2^2)
\right) |\varphi|^2 \right) \D x \D y, \nonumber
\end{eqnarray}
where 
\begin{eqnarray}
c_{22} &=& 3\pi \|\gamma\|_\infty + \pi^2 \|\gamma\|_\infty^2 + 2^{-1} \pi 
\|\gamma'\|_\infty, \\
c_{23} &=& 2^{-1} (\|\gamma\|_\infty + \pi \|\gamma'\|_\infty), \\
c_{24} &=& \pi(1 + 2\|\gamma\|_\infty).  
\end{eqnarray}
By the inequality (\ref{abtoa2b2}), Theorem \ref{hardytheorem} and the fact
that 
\begin{equation}
\chi(x) (\tilde a_1^2(x,y) + \tilde a_2^2(x,y)) \leq \frac {d^2 +
    \pi^2}{\left( \dist (y_0,\partial \Omega_0) \right)^2 \left(x^2 + (y 
    - y_0)^2\right)},
\end{equation}
where $d = \max \supp \gamma$ we obtain
\begin{eqnarray} \label{last}
\form_\beta [\varphi] - \|\varphi\|_{\Ltwo (\Omega_0)}^2
&\geq& \left( \frac 12 - 2 \beta c_{22} \right) \left( \form [\varphi] -
    \|\varphi\|_{\Ltwo (\Omega_0)} ^2 \right) \\
&& \left( \frac {c_{AB}}2 - \beta c_{25}\right) \int_{\Omega_0}
    \frac{|\varphi|^2}{x^2 + (y - y_0)^2} \D x \D y, \nonumber
\end{eqnarray}
where
\begin{eqnarray}
c_{25} &=& (d^2 + \pi^2) \left(2c_{22} + c_{23} + (\dist (y_0, \partial
  \Omega_0))^{-2} (2c_22 +   c_24)\right).
\end{eqnarray}
If we choose
\begin{equation}
\beta_0 = \frac {c_{AB}}{2 c_{25}},
\end{equation}
it follows that the right hand side of \ref{last} is positive.
\end{proof}

\section*{Acknowledgements} 
We would like to thank Timo Weidl for his permanent support and numerous
stimulating discussions throughout the project. Useful comments and
remarks of Denis I. Borisov are also gratefully acknowledged. T.E. has been
partially supported by ESF programme SPECT.


\begin{thebibliography}{99}
%\bibitem[B]{B} M.S.~Birman: The spectrum of singular boundary
%  problems, {\em Translated in Amer. Math. Soc. Trans.} {\bf 53}
%  (1966), 23--80.
%
\bibitem[AHS]{AHS} J.~Avron, J.~Herbst and B.~Simon: Schr\"odinger operators
  with  magnetic fields. I. General interactions, {\em Duke Math. J.} {\bf 45}
  (1978), 847--883.

\bibitem[B]{B} A.A.~Balinsky: Hardy type inequalities for Aharonov-Bohm
  magnetic potentials with multiple singularities, {\em mp\_arc 02-416}. 

\bibitem[BEL]{BEL} A.A.~Balinsky, W.D.~Evans and R.T.~Lewis: On the number of
  negative eigenvalues of Schr\"odinger operators with an Aharonov-Bohm
  magnetic field, {\em mp\_arc 00-405}. 

\bibitem[BS1]{BS2} M.S.~Birman and M.Z.~Solomyak: Spectral theory of
  self-adjoint operators in Hilbert space, {\em D.~Reidel Publishing Company}
  (1987)

\bibitem[BS2]{BS} M.S.~Birman and M.Z.~Solomyak: Schr\"odinger
  Operator. Estimates for number of bound states as
  function-theoretical problem, {\em Amer. Math. Soc. Transl.} (2)
  Vol. {\bf 150}, (1992).

\bibitem[BEGK]{BEGK} D.~Borisov,P.~Exner,R.R.~Gadyl'shin and D.~Krej\v ci\v
  r\'{\i}k: Bound states in weakly deformed strips and layers, {\em Ann. Henri
  Poincar\'e} {\bf 2} (2001), 553--572.

\bibitem[BGRS]{BGRS} W.~Bulla, F.~Gesztesy, W.~Renger and B.~Simon:
  Weakly coupled bound states in quantum waveguides, {\em
  Proc. Amer. Math. Soc.} {\bf 125} (1997),  no. 5, 1487--1495.

\bibitem[DE]{DE} P.~Duclos and P.~Exner: Curvature-induced bound states in
  quantum waveguides in two and three dimensions, {\em Rev. Math. Phys.} {\bf
  7} (1995), 73--102.

\bibitem[EL]{EL} W.D.~Evans and R.T.~Lewis: On the Rellich inequality with
  magnetic potentials, {\em mp\_arc 04-93}.

\bibitem[E\v S]{ES} P.~Exner and P.~\v Seba: Bound states in curved quantum
  waveguides, {\em J.Math. Phys.} {\bf 30} (1989), 2574--2580.

\bibitem[GJ]{GJ} J.~Goldstone and R.L.~Jaffe: Bound states in twisting tubes,
{\em Phys. Rev.} {\bf B45} (1992), 14100--14107.

\bibitem[H]{H} G.~H.~Hardy: Note on a Theorem of Hilbert, {\em Math. Zeit.},
  {\bf 6} (1920), 314--317.

\bibitem[HS]{HS} D.~Hundertmark and B.~Simon: A diamagnetic inequality for
  semigroup differences, {\em mp\_arc 03-78} (2003).

\bibitem[K]{K} T.~Kato: Schr\"odinger operators with singular potentials, {\em
    Israel J. Math.} {\bf 13} (1973), 135--148.

\bibitem[LW]{LW} A.~Laptev and T.~Weidl: Hardy inequalities for
  magnetic Dirichlet forms, {\em Oper. Theory Adv. Appl.} {\bf 108}
  (1999) 299--305.

\bibitem[S]{S} B.~Simon: Maximal and minimal Schr\"odinger forms, {\em
    J. Operator Theory} {\bf 1} (1979), 37--47.

%\bibitem[Y]{Y} D.R.~Yafaev: Scattering matrix for magnetic potentials
%  with couloumb decay at infinity, {\em Integr. equ. oper. theory}
%  {\bf 47} (2003) 217--249.

\end{thebibliography}
\end{document}